\documentclass[reprint,twocolumn,amsmath,amssymb,aps,notitlepage]{revtex4-1}
\usepackage[utf8]{inputenc}
\usepackage{bm}
\usepackage{comment}
\usepackage{braket}
\usepackage{siunitx}
\usepackage{graphicx}
\graphicspath{{img/}}
\usepackage{accents}
\usepackage{stmaryrd}
\usepackage{mathtools}

\newcommand \samcomment[1]{\textcolor{black}{#1}}

\usepackage{soul}
\setstcolor{red}

\begin{document}
\preprint{APS/123-QED} 
\title{Exact Results in Stochastic Processes with Division, Death, and Diffusion:
Spatial Correlations, Marginal Entropy Production, and Macroscopic Currents}
\author{Samuel Cameron}
\author{Elsen Tjhung}
\affiliation{School of Mathematics and Statistics, The Open University, Walton Hall, Milton Keynes MK7 6AA, United Kingdom}
\date{\today}

\begin{abstract}
We consider a generic class of stochastic particle-based models whose state at an instant in time is described by a set of continuous degrees of freedom (e.g. positions), and the length of this set changes stochastically in time due to birth-death processes. 
Using a master equation formalism, we write down the dynamics of the corresponding (infinite) set of probability distributions: this takes the form of coupled Fokker-Planck equations with model-dependent source and sink terms. 
We derive the general expression of entropy production rate for this class of models in terms of path irreversibility. 
To demonstrate the practical use of this framework, we analyze a biologically motivated model incorporating division, death, and diffusion, where spatial correlations arise through the division process.
By systematically integrating out excess degrees of freedom, we obtain the marginal probability distribution, enabling exact calculations of key statistical properties such as average density and correlation functions.
We validate our analytical results through numerical Brownian dynamics simulations, finding excellent agreement between theory and simulation.
Our method thus provides a powerful tool for tackling previously unsolved problems in stochastic birth-death dynamics.
\end{abstract}

\maketitle

\section{Introduction.}

Active matter research has largely focused on systems where energy is locally injected (and subsequently dissipated) to drive the self-propelled motion of individual particles~\cite{Gompper_2020}. 
Synthetic examples include self-propelled colloids with asymmetric surface chemistry~\cite{PhysRevLett.99.048102}, while biological examples range from motile cells and swimming microorganisms to flocks of birds~\cite{RevModPhys.85.1143}.
While most studies focus on motility-driven activity, birth-death processes~\cite{ctx57559520340002316}—common in biological systems—have received less attention. 
These processes play a key role in tissue dynamics, leading to instabilities in otherwise stable glassy systems~\cite{Delarue2016,PhysRevLett.122.208102,PhysRevResearch.2.043334}, bacterial surface proliferation~\cite{doi:10.1143/JPSJ.58.3875,barabasi1995fractal}, and biofilm formation~\cite{doi:10.1137/080739720,Mattei2018,Allen_2019}. 
In particular, cell division introduces spatial correlations along the division axis, yet there remains a lack of probabilistic frameworks capable of accurately tracking these division events in both space and time.

The simplest models of birth and death processes involve only a discrete state space and are often used to describe population dynamics assuming there is
no spatial heterogeneity~\cite{van2011stochastic}.
For example, characterizing the growth of a bacterial colony in terms of the total number of bacteria/cells present can be sufficient if one is only interested in
the global averages, and all physics can be encoded into simple rate constants. 
However, tracking spatial degrees of freedom is often crucial, for example, when simulating the propagating front of a bacterial colony. 
In such cases, the theoretical literature is largely dominated by two approaches:
(i) reaction-diffusion models, which describe coarse-grained particle concentrations at each point in space, incorporating phenomenological terms for division, death, and diffusion~\cite{fisher1937,KPP1937,Allen_2019}; and
(ii) direct particle-based simulations, such as molecular dynamics coupled with stochastic rate equations to track individual trajectories~\cite{Gillespie1977}.

{\color{black}
A third approach is to formulate a master equation using coupled Fokker–Planck equations with source and sink terms.
To track both the number of particles and their spatial positions, the state transitions must allow changes in the number of continuous degrees of freedom and properly account for permutations of particle indices due to indistinguishability~\cite{10.1063/5.0129620}.
Unlike mean field reaction-diffusion models, which focus on effective single-body concentrations in the long-wavelength (hydrodynamic) limit, 
the master equation framework retains the full many-body probability distributions.
However, analytical methods for solving such high dimensional stochastic systems are very limited. 
The most established method is to map the master equation onto a quantum field theory using creation and annihilation operators in a Fock space representation~\cite{PhysRevE.98.062107,Garcia-Millan_2021,Tauber_2014,delRazo2022,refId0,Doi_1976}.
The resulting Doi–Peliti action is generally nonlinear and must be treated perturbatively, typically via renormalization group techniques near the upper critical dimension.

In this paper, we develop a novel method for solving a general master equation in which the number of continuous state variables varies with the discrete state variable.
Our approach is to integrate out the excess continuous degrees of freedom to obtain a hierarchy of marginal probability distributions, analogous to the BBGKY hierarchy in kinetic theory~ \cite{Kardar_2007}.
The time evolution of the one-body marginal distributions, for example, is obtained by integrating the full master equation over all particle positions except that of the first particle (note that the particles are indistinguishable).
This yields a set of one-body distributions that depend on the total number of particles but are functions of only a single spatial coordinate.

As a specific example, we focus on a one-dimensional model of particle division, death, and diffusion (DDD), 
motivated by biological processes such as cell division and apoptosis.
The formalism we present is related to `sizer-timer' models of biological cell division \cite{7801893}, where a collection of cells divides based on their (deterministic) cell size and cell age. A master equation formalism has recently been applied to this class of models \cite{xia_kinetic_2021}, where cell size dynamics were generalised to stochastic dynamics. Phenomenologically, our DDD model is distinct from work in \cite{xia_kinetic_2021} in that division and death is dependent on particle position, and so Brownian motion due to thermal fluctuations is the main stochastic variable. Additionally, we do not consider the age or size of particles in the dynamics (i.e. division and death occurs with time-independent rate constants).
Therefore, our model is} {\color{black} applicable to the dynamics of cells, tissues, and bacterial populations, but may also be relevant to chemical systems where diffusive and reactive timescales are comparable, leading to heterogeneous particle distributions.

More generally, our formalism can describe any birth and death type system where the discrete components are localised in space with coordinates $\{x_1,\dots,x_m\}$. 
Aside from cell-level descriptions, one could also consider a system of receptor molecules in a sea of ligands, 
the latter of which have local concentration gradients. 
Then the locations of unbound receptors, $\{x_1,\dots,x_m\}$, will change due to diffusion and (non-binding) interactions with other receptors,
as well as by binding/unbinding with ligands.
A reaction–diffusion-like equation can then be recovered from the one-body marginal probability distributions.
Thus our formalism provides a direct link between the phenomenological parameters of reaction–diffusion models and the underlying microscopic dynamics.

In the absence of pairwise interactions such as Lennard-Jones potentials, the steady state one-body and two-body distributions can be solved exactly in Fourier space.
From the set of one-body distributions, we can derive the steady state concentration field of the particles, 
which is spatially homogenous in the absence of an external field.
Interestingly, in the presence of an asymmetric ratchet external potential, the system can exhibit a steady state macroscopic current, driven by local particle division.
From the set of two-body distributions, we derive the steady state spatial correlation function, which quantifies the probability of finding a pair of particles separated by a given distance.
Finally, we compare our analytical predictions with particle-based Gillespie simulations, finding excellent agreement. 
This confirms the accuracy and utility of our approach as a powerful analytical tool for studying many-body stochastic systems with particle birth and death.

Finally, we extend the concept of entropy production in stochastic thermodynamics~\cite{Seifert_2012} to our hybrid-state system by evaluating path probabilities in the full high-dimensional phase space.
In addition, we introduce the notion of marginal entropy production, which arises when only a subset of the system’s degrees of freedom is accessible—that is, when the dynamics are described by marginal probability distributions.
To illustrate this, we compute the entropy production rate in our DDD model.
While the total entropy production, evaluated from the full master equation, diverges, the marginal entropy production—obtained by integrating out the excess continuous degrees of freedom—can be finite and even vanish.
}


\section{General formalism}

In this section, we introduce a general formalism to describe a stochastic process in which the number of continuous random variables is itself a random quantity.
Additionally, we extend the concept of entropy production, previously defined only for purely discrete or continuous state systems, 
to our hybrid-state systems.
By integrating out excess degrees of freedom, we show how statistical properties, such as spatial densities and correlations, can be obtained in general.
Finally, we also introduce the notion of marginal entropy production, which quantifies the time-irreversibility of the system when not all degrees of freedom are accessible.

\subsection{Master equation for hybrid-state systems}

At a given time $t$, the state of the system is characterized by a positive integer $m\in\mathbb{Z}^{+}$ and a set of continuous variables 
$\{x\}_m\equiv \{x_1,x_2,\dots,x_m\}$ of length $m$.
Each continuous variable $x_k$ is constrained to a real interval $x_k\in[-L/2,L/2]$, where $L>0$ denotes the system size.
Generalization to higher dimensional space should be straightforward.
As the system evolves in time, the discrete variable $m$ may change stochastically, and consequently, 
the length of the set $\{x\}_m$ may also vary in time.
We refer to such a system as a hybrid-state system.
Physically, $m$ may represent the total number of particles at time $t$ (excluding the vacuum state $m=0$),
while $\{x\}_m$ specifies the positions of these particles (assumed to be indistinguishable).
Due to birth and death processes, we may add or remove particles from the system and thus both $m$ and $\{x\}_m$ are stochastic variables.

We denote $p_m(\{x\}_m,t)d\{x\}_m$ as the probability of having $m$ particles with positions lying within the interval $d\{x\}_m\equiv dx_1\dots dx_m$, centred around $\{x\}_m$, at time $t$.
The probability density is normalized such that
\begin{equation}\label{eq:normalisation}
  \sum_{m=1}^{\infty}\int d\{x\}_m \, p_m(\{x\}_m,t) = 1, \quad\text{for all } t.
\end{equation}
The space integral is implicitly assumed to be taken over the interval $[-L/2,L/2]$ for each coordinate $x_k$.
We define the transition rate (matrix) operator $\hat{W}_{mn}(\{y\}_m\to\{x\}_n;t)$ to be the rate of going from a state of $m$ particles with coordinates $\{y\}_m$ to a state of $n$ particles with coordinates $\{x\}_n$. 
Therefore, $\hat{W}_{mn}$ with $m<n$ represents birth processes, with $n-m$ new particles being generated,
while $\hat{W}_{mn}$ with $m>n$ represents death processes, with $m-n$ particles being killed. 
$\hat{W}_{nn}$ is a transition rate in which particle number does not change, but their positions can change.
We now introduce the hybrid-state master equation, which describes the time evolution of the probability distributions  $\{p_m(\{x\}_m,t)\}$ in terms of these transition rate operators:
\begin{align} \label{eq:hybrid}
      &\partial_t p_n(\{x\}_n,t) =  \nonumber \\
      &\sum_{m=1}^{\infty}\int d\{y\}_m \bigg[  \hat{W}_{mn}(\{y\}_m\to\{x\}_n) p_m(\{y\}_m,t)   \nonumber\\
      &-\hat{W}_{nm}(\{x\}_n\to\{y\}_m) p_n(\{x\}_n,t) \bigg], \quad n\in\mathbb{Z}^+.
\end{align}
Note that $x_k$'s and $y_k$'s are just different labels for the particles' positions along the same coordinate axis.
Eq.~(\ref{eq:hybrid}) evolves from some set of initial conditions $\{p_n(\{x\}_n,0)\}$ and (\samcomment{periodic}) boundary conditions on $x_k$'s to conserve probability. 
From now on, we assume that the transition rate operators $\hat{W}_{mn}(\{y\}_m\to\{x\}_n;t)$ do not depend explicitly on time $t$, 
and thus we have removed the $t$-dependence from $\hat{W}_{mn}$ in Eq.~(\ref{eq:hybrid}).

A particularly interesting situation is when the initial condition is deterministic. 
We define the propagator $G_{mn}(\{x\}_m\to\{y\}_n;\tau,t)$ as the probability that the system is in a state of $n$ particles with coordinates $\{y\}_n$ at time $t+\tau$, given that it was initially in a state of $m$ particles with coordinates $\{x\}_m$ at time $t$.
In other words, $G_{mn}(\{x\}_m\to\{y\}_n;\tau,t)$ is the solution to Eq.~(\ref{eq:hybrid}) with the initial condition
$G(\{x\}_m\to\{y\}_n;0,t) = \delta_{mn}\delta(\{y\}_n-\{x\}_n)$, where $\delta(\{x\}_n-\{y\}_n)\equiv\prod_{k=1}^n\delta(x_k-y_k)$.
To lowest order in $\tau$, the solution to this propagator is found to be
\begin{align}\label{eq:short_time_propagator}
    &G_{mn}(\{x\}_m\to\{y\}_n;\tau) = \delta_{mn}\delta(\{y\}_n-\{x\}_n)  \nonumber\\
    &+\tau\int d\{y'\}_m \, \hat{W}_{mn}(\{y'\}_m\to\{y\}_n)\delta(\{y'\}_m-\{x\}_m),
\end{align}
up to $O(\tau^2)$. 
We have omitted the dependence on absolute time $t$ in the argument of the propagator, since $\hat{W}_{mn}$ does not depend explicitly on time $t$.

\subsection{Entropy production rate for hybrid-state systems \label{subsec:entropyPR}}

The entropy production rate is a measure of the time irreversibility of the stochastic trajectories~\cite{Seifert_2012}.
More specifically, we define a path/trajectory $\bm{X}_M\equiv (\{x(t_0)\}_{n_0},\{x(t_1)\}_{n_1},\dots,\{x(t_M)\}_{n_M})$
as a time-ordered set of points in the phase space; starting at time $t_0$ and visiting state $\{x(t_j)\}_{n_j}$ at successive discrete times $t_j=t_0 + j\tau$ with $j=0,1,\dots,M$. 
$\tau$ is the time interval of successive steps and $\Delta t=M\tau$ the total time passed.
The joint probability of observing this path is
\begin{align}\label{eq:path}
    &\mathcal{P}(\bm{X}_M;t_0,\Delta t)=  \nonumber\\
    &p_{n_0}(\{x(t_0)\}_{n_0},t_0)G_{n_0n_1}(\{x(t_0)\}_{n_0}\to\{x(t_1)\}_{n_1};\tau)\dots  \nonumber\\
    &\times G_{n_{M-1}n_M}(\{x(t_{M-1})\}_{n_{M-1}}\to\{x(t_M)\}_{n_M};\tau),
\end{align}
where $p_{n_0}(\{x(t_0)\}_{n_0},t_0)$ is the probability of observing the system in state $\{x(t_0)\}_{n_0}$ at time $t_0$. 
The reverse path probability is
\begin{equation}\label{eq:reverse_def}
  \mathcal{P}^{\mathcal{R}}(\bm{X}_M;t_0,\Delta t)
  =\mathcal{P}(\bm{X}^{\mathcal{R}}_M;t_0,\Delta t)
\end{equation}
where $\bm{X}^{\mathcal{R}}_M$ is the time-reversed trajectory: $\bm{X}^{\mathcal{R}}_M\equiv (\{x(t_M)\}_{n_M},\{x(t_{M-1})\}_{n_{M-1}},\dots,\{x(t_0)\}_{n_0})$.
We define the entropy change of the hybrid-state system over the time interval $\Delta t$ as the average log-ratio of the forward to backward path probabilities:
\begin{equation}\label{eq:entropy_change}
    \Delta{S}_M = \sum_{\mathbf{X}_M}\mathcal{P}(\bm{X}_M;t_0,\Delta t) 
 	 \ln\left(\frac{\mathcal{P}(\bm{X}_M;t_0,\Delta t)}{\mathcal{P}^\mathcal{R}(\bm{X}_M;t_0,\Delta t)}\right).
\end{equation}
In the equation above, the summation is over all possible trajectories $\bm{X}_M$, visiting every possible state $\{x\}_n$.
By substituting Eq.~(\ref{eq:path}) and (\ref{eq:reverse_def}) into (\ref{eq:entropy_change}), 
then dividing by $\Delta t$ and taking the limit $M\rightarrow\infty$ and $\tau\rightarrow0$, 
we can derive the rate of entropy production $\dot{S}_\text{irr}(t)$ at time $t$ (see Appendix~\ref{app:entropy}):
\begin{align}\label{eq:entropy_production}
    &\dot{S}_\text{irr}(t) = \lim_{\tau\to 0}\frac{1}{\tau}\sum_{m=1}^\infty\sum_{n=1}^\infty\int d\{x\}_m \int d\{y\}_n \nonumber\\
    &\times p_{m}(\{x\}_m,t)G_{mn}(\{x\}_m\to\{y\}_n;\tau)  \nonumber\\
    &\times \ln \left(\frac{p_m(\{x\}_m,t)G_{mn}(\{x\}_m\to\{y\}_n;\tau)}
    			    {p_n(\{y\}_n,t)G_{nm}(\{y\}_n\to\{x\}_m;\tau)}\right),
\end{align}
where the dot represent the total time derivative. 
It can be shown that $\dot{S}_\text{irr}(t)$ is semi-positive definite, i.e., $\dot{S}_\text{irr}(t)\ge0$ (since the summation can be rewritten as a sum of terms in the form $(x-y)\ln(x/y)>0$ when $x>0$ and $y>0$).
Eq.~(\ref{eq:entropy_production}) is the first main result of this paper, 
which generalizes the expression of entropy production previously established for strictly continuous or discrete-state systems~\cite{Gaspard2004,e22111252}.
It accounts for the coupling between the
variable number of particles present in the system and their corresponding position states,
distinguishing it from previous work in which no such coupling was present~\cite{Speck_2018,PhysRevLett.72.1766,Gaspard2007-wo}.
{\color{black} 
A similar expression for entropy production for hybrid continuous and discrete states has been derived in Ref.~\cite{PhysRevE.104.044113}.
However in~\cite{PhysRevE.104.044113}, the number of continuous degrees of freedom, i.e. positions, are assumed to be fixed, whereas in our model they can vary depending on the number of particles.}
Alternatively, $\dot{S}_\text{irr}(t)$ can also defined as the semi-positive definite component of the rate of change of the Shannon entropy.
For our hybrid-state systems, the Shannon entropy is defined to be:
\begin{equation}
  S(t) = -\sum_{n=1}^{\infty}\int d\{x\}_n \, p_n(\{x\}_n,t) \ln \left[p_n(\{x\}_n,t)L^n\right],
\end{equation}
which is an extension of the similar expression given in~\cite{Gaspard2004,e22111252}. {\color{black}Here, $L$ is the system size (units of length) which is required to ensure the argument of the logarithm is dimensionless.}
The rate of change $\dot{S}(t)$ can be decomposed into $\dot{S}(t)=\dot{S}_\text{irr}(t)+\dot{S}_\text{e}(t)$, where $\dot{S}_\text{irr}(t)\ge0$.
In steady state, $\dot{S}(t\rightarrow\infty)$ is zero but $\dot{S}_\text{irr}(t\rightarrow\infty)$ is not zero in general.
Thus, the presence of a finite $\dot{S}_\text{irr}(t\rightarrow\infty)>0$ is a hallmark of a non-equilibrium steady state.

If the operators $\hat{W}_{nm}$ decompose additively into terms associated with translations of particle positions (e.g., diffusion or drift processes) and terms that modify the particle number (i.e., death and birth), then the rate of entropy production similarly separates into two distinct contributions.
In this case, $\dot{S}_\text{irr}$ consists of two terms, one of which {\color{black} is} the expression for entropy production found in discrete-state systems 
(master equation \cite{Gaspard2004}),
while the other {\color{black} is} that of continuous-state systems (Fokker-Planck equation \cite{e22111252}).
Thus, our approach represents a generalization of the entropy production formalism used in the study of, for example, ratchet dynamics in molecular motors \cite{Boksenbojm_2009}.

\begin{figure}
  \includegraphics[width=0.4\textwidth]{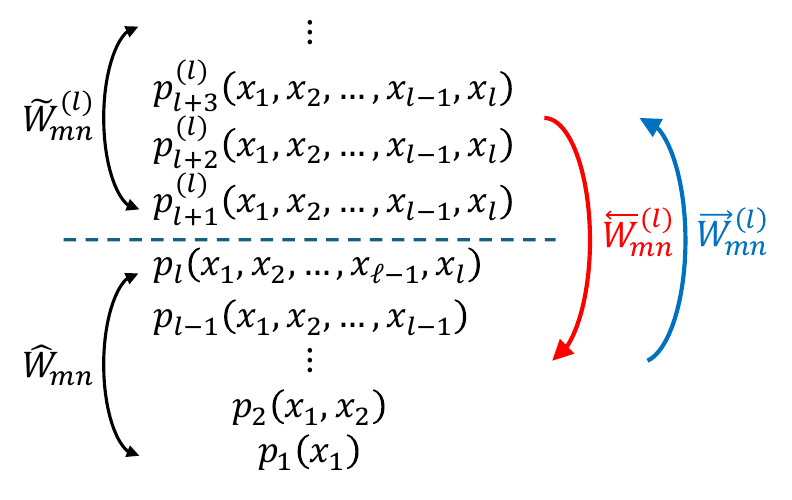}
  \caption{
  The $l$-marginal distribution functions $p_n^{(l)}(x_1,x_2,\dots,x_l)$, where $n>l$, 
  are defined by integrating out the excess coordinates $x_{l+1},x_{l+2},\dots,x_{n}$ from the full distribution functions $p_n(x_1,x_2,\dots,x_n)$.
  The $l$-marginal transition rate operators $W_{mn}^{(l)}(\{y\}_m\rightarrow\{x\}_n)$ can also be defined formally. 
  Here, we distinguish $\overrightarrow{W}_{mn}^{(l)}$ for the transition rate from a state with $m\le l$ particles to a state with $n>l$ particles, 
  and  $\overleftarrow{W}_{mn}^{(l)}$ for the transition rate from a state with $m>l$ particles to a state with $n\le l$ particles.
  Finally, $\widetilde{W}_{mn}^{(l)}$ represents transition rate from a state with $m>l$ particles to a state with $n>l$ particles.
  }\label{fig:marginal}
\end{figure}

\subsection{Marginal probability distributions}

Solving the full probability distribution functions $\{p_n(\{x\}_n,t)\}$  is neither analytically feasible nor particularly useful.
One may instead calculate the $l$-marginal probability distribution functions by integrating out a large proportion of the position variables,
thereby tracking only the first $l$ particle positions:
\begin{equation}\label{eq:marginal}
  p_n^{(l)}(\{x\}_l,t) \equiv
  \int \bigg(\prod_{k={l+1}}^{n} dx_k\bigg) p_n(\{x\}_n,t),
\end{equation}
where the positions of $l$ particles out of $n$ particles are given by $\{x_1,\dots,x_l\}$,
while the positions of the remaining $n-l$ particles are not tracked. 
Note that when $n \leq l$, $p_n^{(l)}(\{x\}_l,t)=p_n(\{x\}_n,t)$, see Fig.~\ref{fig:marginal} for illustration.
This is closely related to the $l$-body density of a system which may be calculated as
\begin{equation}
  \rho^{(l)}(\{x\}_l,t) \equiv \sum_{n=l}^{\infty} \frac{n!}{(n-l)!}p_n^{(l)}(\{x\}_l,t). \label{eq:lbodydensity}
\end{equation}
For example, the average density of a population of particles as a function of space $x$ at time $t$ is given by $\rho(x,t)=\rho^{(1)}(x_1=x,t)$, 
whereas the two-body correlation function is given by $\rho^{(2)}(x_1,x_2;t)$.

We determine the marginal distribution dynamics by integrating the total dynamics, Eq.~(\ref{eq:hybrid}), 
over the excess $n-l$ positions as in Eq.~(\ref{eq:marginal}). 
These integrals can be carried out formally for any set of transition matrices $\hat{W}_{mn}(\{y\}_m\rightarrow\{x\}_n)$.
However, in general, the dynamics of the $l$-marginal distribution $p_n^{(l)}(\{x\}_l,t)$ might depend explicitly
on the $l+1$-marginal distribution $p_n^{(l+1)}(\{x\}_{l+1},t)$. 
This coupling to the $l+1$ marginal is a familiar problem in (constant particle number) statistical mechanics for interacting
systems \cite{Kardar_2007}, where a closure relation such as the ``molecular chaos hypothesis''
\cite{Kardar_2007}
is necessary to write the $l+1$ marginal in terms of the $l$ marginal. 
(Note that in this paper we do not consider explicit interactions between the particles so no closure relation is necessary.)
Once this is done, the $l$-marginal transition rate operators $\hat{W}_{mn}^{(l)}$ can be defined formally.

To this end, it is also necessary to distinguish the $l$-marginal transition rate operators into: $\widetilde{W}_{mn}^{(l)},\overrightarrow{W}_{mn}^{(l)},$ or $\overleftarrow{W}_{mn}^{(l)}$, 
depending on whether they correspond to transitions (i) from a state with $m>l$ to one with $n>l$ particles, (ii) from a state with $m \leq l$ to one with $n > l$ particles, or (iii) vice versa; see Fig.~(\ref{fig:marginal}) for illustration.
This is because constraints must be imposed in two cases:
$m>l$ and $n>l$ [in the definition of $\widetilde{W}_{mn}^{(l)}$, Eq.~(\ref{eq:widetilde_constraint})],
and when $n\leq l < m$  [in the definition of $\overleftarrow{W}_{mn}^{(l)}$, Eq.~(\ref{eq:leftarrow_constraint})]. 
Then, the dynamics for the $l$-marginal distribution, $p_n^{(l)}(\{x\}_l,t)$ can be written as:
\begin{widetext}
\begin{align}\label{eq:formalmarkov_p_n_l_dynamics}
    \partial_t p_n^{(l)}(\{x\}_l,t)
    &=
    \begin{cases}
      \!\begin{aligned}
      & \sum_{m=1}^l\int  d\{y\}_m
      \bigg[\overrightarrow{W}_{mn}^{(l)}(\{y\}_m\to\{x\}_l) p_m(\{y\}_m,t)
      -\overleftarrow{W}_{nm}^{(l)}(\{x\}_l\to\{y\}_m) p_n^{(l)}(\{x\}_l,t) \bigg] \\
      &+\sum_{m=l+1}^{\infty}\int d\{y\}_l
      \bigg[\widetilde{W}_{mn}^{(l)}(\{y\}_l\to\{x\}_l) p_m^{(l)}(\{y\}_l,t)
      -\widetilde{W}_{nm}^{(l)}(\{x\}_l\to\{y\}_l) p_n^{(l)}(\{x\}_l,t) \bigg], && n > l, \\
      & \sum_{m=1}^l\int d\{y\}_m
      \bigg[ \hat{W}_{mn}(\{y\}_m\to\{x\}_n) p_m(\{y\}_m,t)
      -\hat{W}_{nm}(\{x\}_n\to\{y\}_m) p_n(\{x\}_n,t) \bigg] \\
      &+\sum_{m=l+1}^{\infty}\int d\{y\}_l
      \bigg[\overleftarrow{W}_{mn}^{(l)}(\{y\}_l\to\{x\}_n)p_m^{(l)}(\{y\}_l,t)
      -\overrightarrow{W}_{nm}^{(l)}(\{x\}_n\to\{y\}_l)p_n(\{x\}_n,t) \bigg], && n \leq l,
      \end{aligned}
    \end{cases} \\
      &\widetilde{W}_{mn}^{(l)}(\{y\}_l\to\{x\}_l) \equiv \int\bigg(\prod_{k=l+1}^n dx_k\bigg)
       								\hat{W}_{mn}(\{y\}_m\to\{x\}_n),\:\:\: m>l,\: n>l, \label{eq:widetilde_constraint} \\
      &\overleftarrow{W}_{mn}^{(l)}(\{y\}_l\to \{x\}_n) \equiv \hat{W}_{mn}(\{y\}_m\to\{x\}_n), \:\:\: m > l \ge n,  \label{eq:leftarrow_constraint} \\
      &\overrightarrow{W}_{mn}^{(l)}(\{y\}_m\to\{x\}_l) \equiv \int\bigg(\prod_{k=l+1}^ndx_k\bigg)\hat{W}_{mn}(\{y\}_m\to\{x\}_n),\:\:\: n>l\ge m, \label{eq:rightarrow_W}
\end{align}
\begin{equation}  G^{(l)}_{mn}(\{y\}_{m/l}\to\{x\}_{n/l};\tau)=
  \begin{cases}
    G_{mn}(\{y\}_m\to\{x\}_n;\tau),  
    & m \le l, \,\, n \le l, \\
    \tau \int d\{y'\}_l\overleftarrow{W}^{(l)}_{mn}(\{y'\}_l\to\{x\}_n)\delta(\{y'\}_l-\{y\}_l),
    & m > l \ge n, \\
    \delta_{mn}\delta(\{y\}_l-\{x\}_l) + \tau \int d\{y'\}_l
    \widetilde{W}_{mn}^{(l)}(\{y'\}_l\to\{x\}_l)\delta(\{y'\}_l-\{y\}_l),
    & m>l, \,\, n>l, \\
    \tau\int d\{y'\}_m\overrightarrow{W}^{(l)}_{mn}(\{y'\}_m\to\{x\}_l)\delta(\{y'\}_m-\{y\}_m),
    & n > l \ge m.
  \end{cases}  \label{eq:marginal_propagator}
\end{equation}
\end{widetext}
The constraint for $m>l$ and $n>l$ [Eq.~(\ref{eq:widetilde_constraint})] requires $\hat{W}_{mn}$ to be independent of $y_{l+1},y_{l+2},\dots,y_m$,
since the left hand side of the equation is only a function of $y_1,y_2,\dots,y_l$. 
The constraint for $m>l\ge n$ [Eq.~(\ref{eq:leftarrow_constraint})]
requires $\hat{W}_{mn}$ to be independent of $y_{l+1},y_{l+2},\dots,y_m$ for the same reason.
Eq.~(\ref{eq:rightarrow_W}) simply integrates out degrees of freedom, 
so does not constrain the transition matrix operator when $n>l\ge m$, but is defined simply for convenience. 
Thus, Eq.~(\ref{eq:formalmarkov_p_n_l_dynamics}) is \textit{only true} when these constraints on $\hat{W}_{mn}$ are met.
So, subject to Eqs.~(\ref{eq:widetilde_constraint}-\ref{eq:leftarrow_constraint}), 
we find that the marginal distributions can be solved recursively starting from $l=1$ (one-body distributions).

Finally, Eq.~(\ref{eq:marginal_propagator}) defines the short-time marginal propagator subject to Eqs.~(\ref{eq:widetilde_constraint}-\ref{eq:leftarrow_constraint}) up to $O(\tau^2)$. 
The subscript notations $m/l$ and $n/l$ in $G_{mn}^{(l)}$ are meant to identify that there are $m$ particles transitioning to $n$ particles, 
but we emphasize that if $m>l$ and/or $n>l$, the propagator only depends on $\{y\}_l$ and/or $\{x\}_l$ instead of $\{y\}_m$ and/or $\{x\}_n$,
since the last $l-m$ and/or $l-n$ position variables are integrated out. 
Note that when $n\leq l$ and $m \leq l$, this marginal propagator just reduces to Eq.~(\ref{eq:short_time_propagator}) as it should.

\subsection{Marginal entropy production rates}

In addition to the marginal probability distributions, one can also construct an analogue to the entropy production rate Eq.~(\ref{eq:entropy_production}), considering up to the first $l$ position variables. 
We refer to this quantity as the $l$-marginal entropy production rate, and denote it by $\dot{S}^{(l)}_\text{irr}$. 
Similar to the total entropy production rate derived in Sec.~\ref{subsec:entropyPR} and Appendix~\ref{app:entropy},
$\dot{S}^{(l)}_\text{irr}$ is defined as the expectation of the logarithm of the ratio of forward to reverse path probabilities in the reduced phase space. 
In our derivation, we require that Eqs.~(\ref{eq:widetilde_constraint}-\ref{eq:leftarrow_constraint}) are satisfied. 
Our $l$-marginal entropy production rate is related to, yet distinct from, 
the \emph{coarse-grained} entropy production rate obtained by spatially averaging the particle current $j(x)\rightarrow\left<j\right>(x)$~\cite{PhysRevE.105.L042601,PhysRevE.107.024122,Agranov_2024}, similar to a renormalization group procedure~\cite{PhysRevLett.124.240604}.

Following the procedure for the full entropy production rate calculation described in Sec.~\ref{subsec:entropyPR}, 
one can define the $l$-marginal path probability
\begin{equation}\label{eq:marginal_path}
  \begin{split}
    &\mathcal{P}^{(l)}(\bm{X}_M;t_0,\Delta t)=p_{n_0}^{(l)}(\{x(t_0)\}_{n_0/l},t_0)\\
    &\times G^{(l)}_{n_0 n_1}(\{x(t_0)\}_{n_0/l}\to\{x(t_1)\}_{n_1/l};\tau)\dots\\
    &\times G^{(l)}_{n_{M-1}n_M}(\{x(t_{M-1)}\}_{n_{M-1}/l}\to\{x(t_M)\}_{n_M/l};\tau)
  \end{split}
\end{equation}
where the propagators $G^{(l)}_{nm}$ (and the initial probability $p_{n_0}^{(l)}$) depend at most on the first $l$ positions as mentioned below Eq.~(\ref{eq:marginal_propagator}). 
Additionally, the $l$-marginal reverse path probability is defined as in Eq.~(\ref{eq:reverse_def}), but using Eq.~(\ref{eq:marginal_path}) instead of Eq.~(\ref{eq:path}).
By a very similar calculation to the full entropy production rate, 
we arrive at the positive definite, $l$-marginal entropy production rate, which the second main result of our paper, 
\begin{widetext}
\begin{align}
    \dot{S}^{(l)}_\text{irr}(t)
    &= \lim_{\tau\to0}\Bigg\{\frac{1}{\tau}\sum_{n=1}^l\sum_{m=1}^l\int d\{x\}_n \int d\{y\}_m \, G_{mn}(\{y\}_m\to\{x\}_n;\tau)p_m(\{y\}_m)
    	\ln\left[\frac{G_{mn}(\{y\}_m\to\{x\}_n;\tau)p_m(\{y\}_m)}{G_{nm}(\{x\}_n\to\{y\}_m;\tau)p_n(\{x\}_n)}\right] \nonumber\\
    &+ \frac{1}{2\tau}\sum_{n=1}^l\sum_{m=l+1}^{\infty}\int d\{x\}_n\int d\{y\}_l \, \overleftarrow{G}^{(l)}_{mn}(\{y\}_l\to\{x\}_n;\tau)p_m^{(l)}(\{y\}_l) 
    	\ln\left[\frac{\overleftarrow{G}^{(l)}_{mn}(\{y\}_l\to\{x\}_n;\tau)p_m^{(l)}(\{y\}_m)}{\overrightarrow{G}^{(l)}_{nm}(\{x\}_n\to\{y\}_l;\tau)p_n(\{x\}_n)} \right] \nonumber\\
    &+\frac{1}{\tau}\sum_{n=l+1}^{\infty}\sum_{m=l+1}^{\infty}\int d\{x\}_l\int d\{y\}_l \, \widetilde{G}^{(l)}_{mn}(\{y\}_l\to\{x\}_l;\tau)p_m^{(l)}(\{y\}_l) 
    	\ln\left[\frac{\widetilde{G}^{(l)}_{mn}(\{y\}_l\to\{x\}_l;\tau)p_m^{(l)}(\{y\}_m)}{\widetilde{G}^{(l)}_{nm}(\{x\}_l\to\{y\}_l;\tau)p_n^{(l)}(\{x\}_n)}\right] 
    \Bigg\}, \label{eq:S-marginal}
\end{align}
\end{widetext}
where $\overleftarrow{G}_{mn}$, $\overrightarrow{G}_{mn}$ and $\widetilde{G}_{mn}$ are the propagators associated with $\overleftarrow{W}_{mn}$, $\overrightarrow{W}_{mn}$, and $\widetilde{W}_{mn}$ respectively.
Additionally, the $l$-marginal Shannon entropy can be similarly defined,
\begin{align}
 S^{(l)}(t) =& -\sum_{n=1}^l\int d\{x\}_n \, p_n(\{x\}_n,t)
    \ln\big[p_n(\{x\}_n,t)L^n\big] \nonumber\\
    -\sum_{n=l+1}^{\infty}&\int d\{x\}_l \, p_n^{(l)}(\{x\}_l,t)
    \ln\big[p_n^{(l)}(\{x\}_l,t)L^l\big], \label{eq:coarse_grained_ROEC}
\end{align}
and the marginal rate of entropy production rate $\dot{S}_\text{irr}^{(l)}(t)$ can also be obtained from the semi-positive definite part of $\dot{S}^{(l)}(t)$.

\section{Division, death, and diffusion in the absence of external field \label{sec:DDD}}

Having presented a formal description of stochastic hybrid-state systems and their (marginal) probability distributions and entropy production rates, we next proceed to apply these results to a specific model, which we refer to as the division, death, and diffusion (DDD) model.
In this model, particles can divide into two with rate $r_b$, die with rate $r_a(n-1)$, or undergo random walks with diffusion constant $D$, where 
$n$ denotes the total number of particles at a given time (see Fig.~\ref{fig:DDD}). 
The death rate scales with $n-1$ to prevent transitions into the vacuum state ($n=0$).
In stochastic thermodynamics, the rate constants $r_a$ and $r_b$  are typically expressed as exponentials of activation energies divided by temperature~\cite{Pietzonka_2018,Seifert_2012}. 
Although particles do not interact directly, division events create spatial correlations between the resulting particles [see Fig.~\ref{fig:DDD}(a)].
We will explore how non-equilibrium processes, such as particle division and death, influence the entropy production rate. 
By first deriving the $1$-marginal and $2$-marginal probability distributions, we will also demonstrate how the average density and two-body correlations can be computed exactly.

\subsection{Perfectly homogeneous system (division and death only)\label{subsec:nodiffusion}}

As a warm-up to our later calculations, we first present the entropy production rate for a division and death system which neglects spatial fluctuations completely. 
In this case, the state of the system is characterized entirely by the number of particles $n$ at a given time $t$ with the corresponding probability mass function $p_n(t)$. 
The time evolution for $p_n(t)$ is exactly described by the autocatalytic process
\footnote{See section 6.9 in \cite{van2011stochastic}.}
\begin{equation}\label{eq:homo-master}
  \begin{split}
    \frac{d p_n(t)}{dt} = & \underbrace{r_an(n+1)}_{W_{n+1,n}} p_{n+1}(t) + \underbrace{r_b(n-1)}_{W_{n-1,n}}p_{n-1}(t)  \\
    & \underbrace{- \left[r_an(n-1) + r_bn \right]}_{W_{nn}}p_n(t),
  \end{split}
\end{equation}
from which we can read off the transition rates:
\begin{align}
    W_{mn} &= r_a n(n+1) \delta_{m,n+1} + r_b(n-1) \delta_{m,n-1} \nonumber\\
    &- \left[r_an(n-1) + r_b n\right]\delta_{m,n}.
\end{align}
Note that the transition rate only increments the total number of particles by $\pm1$ at most.
Using \samcomment{the} generating function \samcomment{formalism}~\cite{van2011stochastic}, Eq.~(\ref{eq:homo-master}) can be solved exactly in steady state:
\begin{equation} 
  \pi_n\equiv \lim_{t\to\infty} p_n(t) = \bigg(\frac{r_b}{r_a}\bigg)^n\frac{1}{n!}\frac{1}{e^{r_b/r_a}-1}. \label{eq:pi_n}
\end{equation}
Consequently, the average number of particles $\left<n\right>$ and the variance in the number of particles $\left<n^2\right>-\left<n\right>^2$ can also be computed in steady state:
\begin{align} 
    \left<n\right>_\text{ss} &= \sum_{n=1}^\infty n \pi_n = \left(\frac{r_b}{r_a}\right) \frac{e^{r_b/r_a}}{e^{r_b/r_a}-1} \label{eq:n-avg} \\
    \left<n^2\right>_\text{ss}-\left<n\right>^2_\text{ss} &= \frac{r_be^{r_b/r_a}\left( e^{r_b/r_a} -1 - r_b/r_a \right)}{r_a(e^{r_b/r_a}-1)^2},
\end{align}
where the subscript `ss' indicates averaging in the steady state.
When $r_b\gg r_a$, the average number of particles in the system is given by $\langle n \rangle_\text{ss} \approx r_b/r_a$ in steady-state. 

Direct substitution of this steady-state mass function, $\pi_{n}$, along with the transition rate, $W_{mn}$, into the definition of the entropy production rate for discrete state systems~\cite{Gaspard2004,e22111252} yields, in steady state,
\begin{equation}\label{eq:homo_entropy}
  \lim_{t\to\infty}\dot{S}_\text{irr}(t)=\sum_{m,n}\pi_nW_{nm}\ln\left(\frac{\pi_n W_{nm}}{\pi_m W_{mn}}\right) = 0.
\end{equation}
Therefore, there is no entropy production in the spatially homogeneous case. 
We note that the assumption of homogeneously distributed particles is often employed when using master equations to describe chemical reactions, 
given the relatively slow time-scale at which reactions occur relative to the spatial diffusion of reacting molecules.

\subsection{Spatial diffusion included (DDD model)}

\begin{figure}
  \def\svgwidth{0.5\textwidth}
  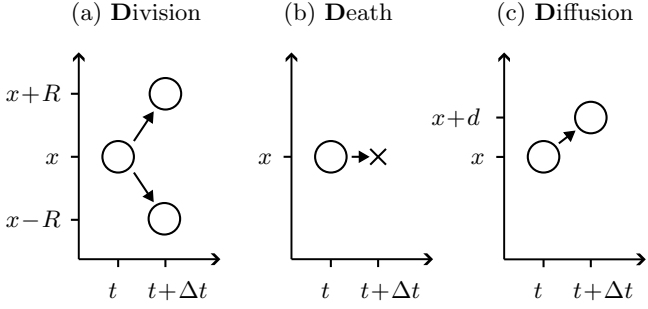
  \vspace{0pt}
  \caption{Schematic of (a) division, (b) death, and (c) diffusion processes in the DDD model. 
    Exactly one of these processes occurs in timestep $\Delta t$ for each particle (in the limit $\Delta t\to 0$).
    Here we highlight three possible transitions of a single particle which starts at position $x$ and time $t$. 
    Division/birth occurs with rate $r_b$ and death with rate $r_a(n-1)$. 
    Unbiased diffusion occurs at every time step where division or death does not occur.
    For every division event, the parent particle at $x$ is removed and two new particles are created at $x+R$ and $x-R$.
    We also assume periodic boundary conditions at $x=-L/2$ and $x=L/2$, and $x\in[-L/2,L/2]$.}\label{fig:DDD}
\end{figure}

We now extend the previous analysis to a one-dimensional system of particles that can undergo division and death processes while diffusing independently under periodic boundary conditions with a period length $L$; see Fig.~\ref{fig:DDD}.
Although the particles do not interact directly, each division event generates spatial correlations, as division produces two new particles separated by a fixed distance $2R$, see Fig.~\ref{fig:DDD}(a). {\color{black} Therefore, the DDD model is an extension of the simple Branching Brownian Motion processes \cite{skorokhod_branching_1964} by including a non-zero splitting radius $R$.}
We also assume the absence of an external field. In Sec.~\ref{sec:geoproj}, we will explore the effect of an external field, in particular a ratchet potential.
The master equation for the full dynamics is given by:
\begin{widetext} 
\begin{align} \label{eq:hetero-master}
    &\partial_tp_n(\{x\}_n,t) = \sum_{i=1}^n -\partial_{x_i} \left[ \hat{j}_{x_i} p_n(\{x\}_n,t) \right]
    						   + r_a n(n+1) \int dx_{n+1}\,p_n(x_1,\dots,x_n,x_{n+1}) - \left[ r_an(n-1) + bn \right] p_n(\{x\}_n,t) \nonumber\\
	& +\frac{r_b}{n}\sum_{i=1}^n \sum_{j=1}^{i-1} \int dy\, \delta(x_i - f_{-R}(y))\delta(x_j - f_{+R}(y)) 
										  p_{n-1}(x_1,\dots,x_{j-1},x_{j+1},\dots,x_{i-1},y,x_{i+1},\dots,x_n;t) \nonumber\\
	& +\frac{r_b}{n}\sum_{i=1}^n \sum_{j=i+1}^{n} \int dy\, \delta(x_i - f_{-R}(y))\delta(x_j - f_{+R}(y)) 
										  p_{n-1}(x_1,\dots,x_{i-1},y,x_{i+1},\dots,x_{j-1},x_{j+1},\dots,x_n;t),
\end{align}
\end{widetext}
where we have introduced the operator $\hat{j}_{x_i}$ and the function $f_{\pm R}(y)$ to be:
\begin{equation}
\hat{j}_{x_i} = -D\partial_{x_i} \quad\text{and}\quad f_{\pm R}(y) = y\pm R. \label{eq:j-f}
\end{equation}
In Sec.~\ref{sec:geoproj}, we will consider a generalization to $\hat{j}_{x_i}$ and $f_{\pm R}(y)$.
The first term in Eq.~(\ref{eq:hetero-master}) represents diffusion.
The second term accounts a death event occurring at position $x_{n+1}$.
The third term acts as a sink term for the $n$-particle state, capturing the loss of probability due to division or death events of any particle.
The fourth and the fifth terms describe a division event at position $y$, where a particle splits into two new particles located at $x_i=y-R$ and $x_j=y+R$.
The summation ensures permutation over all position indices, maintaining the invariance of the probability distribution $p_n(x_1,\dots,x_n)$ under the exchange of any two particle indices, as required by the indistinguishability condition.

A more realistic variation of the model incorporates stochasticity in the division separation distance $2R$. 
Instead of the two new particles being created at a fixed separation $\pm R$ relative to the parent particle, 
their separation is given by $\pm r$, where $r$ is a random variable drawn from a Gaussian distribution with mean $R$ and variance $\sigma_R^2$.
In this case, the Dirac delta function $\delta(x)$ in Eq.~(\ref{eq:hetero-master}) is replaced by a Gaussian function:
\begin{equation} 
  \delta_{\sigma_R}(x) = \frac{1}{\sqrt{2\pi\sigma_R^2}}e^{-\frac{x^2}{2\sigma_R^2}}. \label{eq:Gaussian}
\end{equation}
The deterministic division separation can then be recovered by taking the limit $\sigma_R\rightarrow 0$.
In our simulation results, we assume a deterministic division separation of $2R$ for simplicity.
However, in Sec.~\ref{sec:1-marginal} we also discuss the effect of random separation distance on the rate of entropy production.

To get the $0$-marginal probability distributions $p_n^{(0)}(t)$, we integrate Eq.~(\ref{eq:hetero-master}) over all position variables $x_1,x_2,\dots$.
The result is identical to the perfectly homogenous master equation (\ref{eq:homo-master}) with the identification $p_n^{(0)}(t)=p_n(t)$.
Thus at this level, the $0$-marginal entropy production rate $\dot{S}_\text{irr}^{(0)}$ is zero in steady state.

\subsection{$1$-marginal distributions and $1$-marginal entropy production rate \label{sec:1-marginal}} 

Integrating over $x_2,x_3,\dots$ in Eq.~(\ref{eq:hetero-master}) gives us the time evolution for the $1$-marginal probability distributions
\begin{align}
    &\partial_tp_n^{(1)}(x_1) = D\partial_{x_1}^2p_n^{(1)}(x_1) + r_an(n+1) p_{n+1}^{(1)}(x_1) \nonumber\\
    & - \left[ r_an(n-1) + r_b n \right] p_n^{(1)}(x_1) + \frac{r_b}{n}(n-1)(n-2)p_{n-1}^{(1)}(x_1) \nonumber\\
    & + \frac{r_b}{n}(n-1) \left[ p_{n-1}^{(1)}(x_1-R) + p_{n-1}^{(1)}(x_1+R)  \right],  \label{eq:1-marginal}
\end{align}
for $n\in\mathbb{Z}^+$.
Note that we have not written the time dependence explicitly inside the argument of $p_n^{(1)}$ for brevity.
From Eq.~(\ref{eq:1-marginal}), we can then read off the $1$-marginal transition rates:
\begin{align}
    &\widetilde{W}^{(1)}_{n-1,n}(y_1\rightarrow x_1) =  \frac{r_b(n-1)(n-2)}{n} \delta(y_1-x_1) \nonumber\\
    								    & \quad + \frac{r_b(n-1)}{n}\left[\delta(y_1-x_1+R) + \delta(y_1-x_1-R)\right] \label{eq:W-1-1}\\
    &\widetilde{W}^{(1)}_{n,n}(y_1\rightarrow x_1) = D\partial_{x_1'}^2\delta(y_1-x_1) \\
    &\widetilde{W}^{(1)}_{n+1,n}(y_1\rightarrow x_1) = r_a n(n-1)\delta(y_1-x_1), \label{eq:W-1-3}
\end{align}
and $\hat{W}_{11}=\widetilde{W}^{(1)}_{11}$, $\overleftarrow{W}_{21}^{(1)}=\widetilde{W}^{(1)}_{21}$, and $\overrightarrow{W}^{(1)}_{12}=\widetilde{W}^{(1)}_{12}$.
Thus for $1$-marginal distributions, it turns out that we do not need to distinguish between states with $n\le l$ particles and those with $n> l$ particles.

Since $x_1\in[-L/2,L/2]$ and assuming periodic boundary conditions, we can Fourier transform $p_n^{(1)}(x_1,t)$:
\begin{equation} 
p_n^{(1)}(x_1,t) = \sum_{k\in\mathbb{Z}} c_n^{(1)}(k,t) e^{i\frac{2\pi k}{L}x_1}. \label{eq:p-1-FT}
\end{equation}
Eq.~(\ref{eq:1-marginal}) can then be Fourier transformed to get
\begin{equation} 
\partial_t \bm{c}^{(1)}(k,t) = \bm{A}^{(1)}(k) \bm{c}^{(1)}(k,t),
\end{equation}
where $k\in\mathbb{Z}$.
In the above equation, $\bm{c}^{(1)}(k,t)=[c_1^{(1)}(k,t),c_2^{(1)}(k,t),\dots]^T$ is a semi-infinite column vector and $\bm{A}^{(1)}(k)$ is a semi-infinite tridiagonal matrix with non-zero elements given by
  \begin{align}
    A_{n,n-1}^{(1)}(k) &= \frac{r_b(n-1)}{2 n} \left[ \cos\left(\frac{2\pi Rk}{L}\right) + \frac{n-2}{2}  \right], \label{eq:A-1-1} \\
    A_{n,n}^{(1)}(k) &= -\frac{D(2\pi k)^2}{L^2} - [r_an(n-1) + r_bn], \\
    A_{n,n+1}^{(1)}(k) &= r_a n(n+1). \label{eq:A-1-3}
\end{align}
{\color{black}
From a probabilistic argument, the real parts of all eigenvalues of $\bm{A}(k)$ must be non-positive for all values of $k$.
Otherwise, the probability distribution would diverge over time instead of remaining normalized to unity.
Thus, the steady state solution corresponds to the eigenvector associated with the zero eigenvalue, which happens to occur at $k=0$.}
The $k=0$ mode represents a spatially homogenous solution, where $p_n^{(1)}(x_1)$ is independent of the position $x_1$ for all $n\in\mathbb{Z}^+$.
Moreover, the steady state $1$-marginal distribution matches the perfectly homogenous solution (apart from the normalization factor $1/L$): $p_n^{(1)}(x_1,\infty)=\pi_n/L$, where $\pi_n$ is given in Eq.~(\ref{eq:pi_n}) in Sec.~\ref{subsec:nodiffusion}.

Correspondingly, the average density of the particles can be computed using Eq.~(\ref{eq:lbodydensity}), and in steady state, this is given by
\begin{equation}
\rho(x,\infty) = \lim_{t\rightarrow\infty }\sum_{n=1}^\infty n p_n^{(1)}(x,t) = \sum_{n=1}^\infty n \frac{\pi_n}{L} = \frac{\left<n\right>_\text{ss}}{L},
\end{equation}
where $\left<n\right>_\text{ss}$ is the average total number of particles in steady state, given in Eq.~(\ref{eq:n-avg}). 
Thus, while the particle density $\rho(x,t)$ may initially exhibit spatial heterogeneities, over time, these spatial variations gradually diminish. 
In the steady state, the average density becomes spatially uniform, $\rho(x,\infty)=$ constant, as expected due to translational symmetry in the system.

We can now substitute the steady state distributions $\pi_n/L$ and the transition rates $\widetilde{W}^{(1)}_{mn}$  [Eqs.~(\ref{eq:W-1-1}-\ref{eq:W-1-3})] into  Eq.~(\ref{eq:S-marginal}) to get the $1$-marginal entropy production rate:
\begin{align}
    &\dot{S}_\text{irr}^{(1)} = \sum_{n=2}^{\infty}\bigg(\frac{r_b}{r_a}\bigg)^n\frac{r_a}{n!L}\frac{(n-1)}{e^{r_b/r_a}-1} \nonumber\\
    &\times\int dx \int dy \left[\delta(x+R-y) + \delta(x-R-y) - 2\delta(x-y)\right] \nonumber\\
    &\times \ln\left[\frac{(n-2)\delta(x-y) + \delta(x+R-y) + \delta(x-R-y)}{n\delta(x-y)}\right]. \label{eq:S-1}
\end{align}
To compute the integral of a logarithm of a delta function in Eq.~(\ref{eq:S-1}),
we first need to regularize the delta-function as a Gaussian function $\delta_{\sigma_R}(x)$, see Eq.~(\ref{eq:Gaussian}).
Physically, we now allow the division separation distance to be random, rather than fixed at $\pm R$, 
and $\sigma_R^2$ represents the variance of the separation distance, see discussion above Eq.~(\ref{eq:Gaussian}).
The $1$-marginal entropy production rate can then be computed exactly
\begin{equation}
  \dot{S}_\text{irr}^{(1)} =
  \begin{cases}
    \infty,  & R\neq 0 \,\,\&\,\, \sigma_R\rightarrow 0, \\
    \frac{R^2}{\sigma_R^2}\left(\frac{r_b}{e^{r_b/r_a}-1}+r_b-r_a \right), & R\neq 0 \,\,\&\,\, \sigma_R>0, \\
    0, & R=0.
  \end{cases} \label{eq:1m_epr}
\end{equation}
Thus, even though the steady state average density $\rho(x,t\rightarrow\infty)$ is spatially uniform, 
we still have a positive and finite rate of entropy production in the steady state (for the case of $R\neq0$ and $\sigma_R>0$).
This is because entropy production captures the dynamics between different particle number states, e.g., between $p_n^{(1)}(x_1)$ and $p_{n+1}^{(1)}(x_1)$.
In the limit of deterministic separation distance $\sigma_R\rightarrow0$, the rate of entropy production becomes infinite.
This infinite entropy production stems from the improbability of the reverse process: two particles approaching each other to a separation distance $2R$, being removed, and replaced by a single particle at their centre of mass.

\subsection{$2$-marginal distributions and $2$-body correlation function}

In this section, we assume a fixed division separation of $2R$ after each division event, 
corresponding to the limit $\sigma_R\rightarrow0$, to make our calculations analytically tractable.
Integrating over $x_3,x_4,x_5,\dots$ in Eq.~(\ref{eq:hetero-master}), we obtain the time evolution for the $2$-marginal probability distributions:
\begin{align}
&\partial_t p_n^{(2)}(x_1,x_2) \nonumber\\
&= \left\{D(\partial_{x_1}^2+\partial_{x_2}^2) - [r_an(n-1) + r_b n]\right\}p_n^{(2)}(x_1,x_2)  \nonumber\\
&+ r_an(n+1)p_{n+1}^{(2)}(x_1,x_2)  + \frac{r_b}{n}(n-2)(n-3) p_{n-1}^{(2)}(x_1,x_2)  \nonumber\\
&+ \frac{r_b}{n} \left[ p_{n-1}^{(2)}(x_1+R,x_2) + p_{n-1}^{(2)}(x_1-R,x_2) \right] \nonumber\\
&+ \frac{r_b}{n} \left[ p_{n-1}^{(2)}(x_1,x_2+R) + p_{n-1}^{(2)}(x_1,x_2-R) \right] \nonumber\\
&+ \frac{r_b}{n} \delta(x_2-x_1-2R)p_{n-1}^{(1)}(x_1+R) \nonumber\\
&+ \frac{r_b}{n} \delta(x_2-x_1+2R)p_{n-1}^{(1)}(x_1-R), \label{eq:2-marginal-1}
\end{align}
for $n\ge 2$, and,
\begin{align}
\partial_t p_1(x_1) = D\partial_{x_1}^2 p_1(x_1) + 2r_ap_2^{(1)}(x_1) - r_b p_1(x_1). \label{eq:2-marginal-2}
\end{align}
for $n=1$. 
Notice that the time evolution of the $2$-marginal distributions depends on the $1$-marginal distributions.
Therefore, solving Eqs.~(\ref{eq:2-marginal-1}-\ref{eq:2-marginal-2}) also requires solving Eq.~(\ref{eq:1-marginal}) simultaneously.
To solve these equations, we again use the Fourier transform (assuming periodic boundary conditions):
\begin{equation}
p_n^{(2)}(x_1,x_2;t) = \sum_{k_1,k_2\in\mathbb{Z}} c_n^{(2)}(k_1,k_2;t) e^{i\frac{2\pi k_1}{L}x_1}e^{i\frac{2\pi k_2}{L}x_2}, \label{eq:p-2-FT}
\end{equation}
while the Fourier transform of $p_n^{(1)}(x_1,t)$ is still given by Eq.~(\ref{eq:p-1-FT}).
Using Eqs.~(\ref{eq:p-1-FT}) and (\ref{eq:p-2-FT}), Eq.~(\ref{eq:2-marginal-1}) can then be Fourier transformed to:
\begin{align}
&\partial_t c_n^{(2)}(k_1,k_2;t) \nonumber \\
&= \left\{ -\frac{D(2\pi)^2(k_1^2+k_2^2)}{L^2} - [r_an(n-1) + r_b n] \right\} c_n^{(2)} \nonumber \\
&+ r_an(n+1) c_{n+1}^{(2)}+ \frac{r_b}{n}(n-2)(n-3) c_{n-1}^{(2)} \nonumber \\
&+ \frac{2r_b}{n}(n-2)\left[\cos\left(\frac{2\pi k_1R}{L}\right) + \cos\left(\frac{2\pi k_2R}{L}\right)\right] c_{n-1}^{(2)} \nonumber \\
&+ \frac{2r_b}{Ln} \cos\left[\frac{2\pi R}{L}(k_1-k_2)\right] c_{n-1}^{(1)}(k_1+k_2;t),
\end{align}
where $n\ge2$ and the argument of $c_n^{(2)}$ on the right hand side of the equation above is implied to be $(k_1,k_2;t)$.
The time evolution for $c_n^{(1)}(k,t)$ and $c_n^{(2)}(k_1,k_2;t)$ can then be written as a matrix form:
\begin{equation}
\partial_t \bm{c}^{(1)+(2)}(k_1,k_2;t) = \bm{A}^{(1)+(2)}(k_1,k_2)\bm{c}^{(1)+(2)}(k_1,k_2;t), \label{eq:c-1-2-dot}
\end{equation}
where $k_1$ and $k_2$ are integers.
In the above equation, $\bm{c}^{(1)+(2)}(k_1,k_2;t)$ is the column vector
\begin{equation}
\bm{c}^{(1)+(2)}(k_1,k_2;t) = 
\begin{bmatrix}
c_1^{(1)}(k_1+k_2;t) \\
c_2^{(1)}(k_1+k_2;t) \\
\vdots \\
c_2^{(2)}(k_1,k_2;t) \\
c_3^{(2)}(k_1,k_2;t) \\
\vdots  
\end{bmatrix}
\end{equation}
and $\bm{A}^{(1)+(2)}(k_1,k_2)$ is the block matrix
\begin{equation}
\bm{A}^{(1)+(2)}(k_1,k_2) =
\begin{bmatrix}
\bm{A}^{(1)}(k_1+k_2) & \bm{0} \\
\bm{B}(k_1-k_2)  & \bm{A}^{(2)}(k_1,k_2).
\end{bmatrix}  \label{eq:A-1-2}
\end{equation}
In the definition above, $\bm{A}^{(1)}(k)$ is a semi-infinite tridiagonal square matrix, as defined in Eqs.~(\ref{eq:A-1-1}-\ref{eq:A-1-3}),
$\bm{B}(k)$ is a semi-infinite diagonal matrix, with its diagonal elements given by
\begin{equation}
B_{n-1,n-1}(k) = \frac{2r_b}{nL} \cos\left(\frac{2\pi k R}{L}\right), 
\end{equation}
and finally $\bm{A}^{(2)}(k_1,k_2)$ is another semi-infinite tridiagonal square matrix, with its non-zero elements given by
\begin{align}
A_{n-1,n-2}^{(2)}(k_1,k_2) &= \frac{r_b(n-2)}{n}\bigg[(n-3) + 2\cos\left(\frac{2\pi k_1R}{L}\right) \nonumber\\
			 & + 2\cos\left(\frac{2\pi k_2R}{L}\right) \bigg]  \\
A_{n-1,n-1}^{(2)}(k_1,k_2) &= -\frac{D(2\pi)^2(k_1^2+k_2^2)}{L^2} \nonumber\\
					  &- \left[k_an(n-1) + k_bn\right]  \\
A_{n-1,n}^{(2)}(k_1,k_2) &= r_an(n+1).
\end{align}
$\bm{0}$ in Eq.~(\ref{eq:A-1-2}) is a semi-infinite matrix with zero as its elements.

The matrix equation (\ref{eq:c-1-2-dot}) cannot be solved directly.
Instead, we first need to truncate the semi-infinite matrix $\bm{A}^{(1)}$ to obtain a finite $N\times N$ matrix, denoted as $\bm{A}^{(1)}_{N\times N}$, 
which consists of the first $N$ rows and $N$ columns of the original matrix $\bm{A}^{(1)}$.
This approximation is valid as long as $N$ is much larger than the average number of particles, i.e. $N\gg\left<n\right>_\text{ss}$.
Similarly, we also truncate  $\bm{A}^{(2)}$ to obtain $\bm{A}^{(2)}_{(N-1)\times(N-1)}$,
and truncate $\bm{B}$ to obtain $\bm{B}_{(N-1)\times N}$.
The matrix $\bm{A}^{(1)+(2)}$ in Eq.~(\ref{eq:A-1-2}) can then be truncated as a $(2N-1)\times(2N-1)$ finite block matrix:
\begin{equation}
\bm{A}^{(1)+(2)}_{(2N-1)\times(2N-1)} = 
\begin{bmatrix}
\bm{A}^{(1)}_{N\times N}& \bm{0}_{N\times(N-1)} \\
\bm{B}_{(N-1)\times N}  & \bm{A}^{(2)}_{(N-1)\times(N-1)}
\end{bmatrix},  \label{eq:A-1-2-N}
\end{equation}
and we have $\bm{A}^{(1)+(2)}=\lim_{N\rightarrow\infty}\bm{A}^{(1)+(2)}_{(2N-1)\times(2N-1)}$.
We are generally interested in the steady state solution, i.e., we are looking for the zero eigenvalue of the matrix $\bm{A}^{(1)+(2)}_{(2N-1)\times(2N-1)}$.
To achieve this, we shall use the following result.

\emph{Proposition} (Eigenvalues of block matrices). If $\lambda$ is an eigenvalue of $\bm{A}^{(1)}_{N\times N}$ or $\bm{A}^{(2)}_{(N-1)\times (N-1)}$, then $\lambda$ is also an eigenvalue of $\bm{A}^{(1)+(2)}_{(2N-1)\times(2N-1)}$. 
The proof follows directly from the properties of the determinant of a block matrix~\cite{Silvester_2000}.

\begin{figure}
  \def\prefix {../experiments/2024_11_16/two_marginal_densities_convergence/}
  \includegraphics[width=0.5\textwidth]{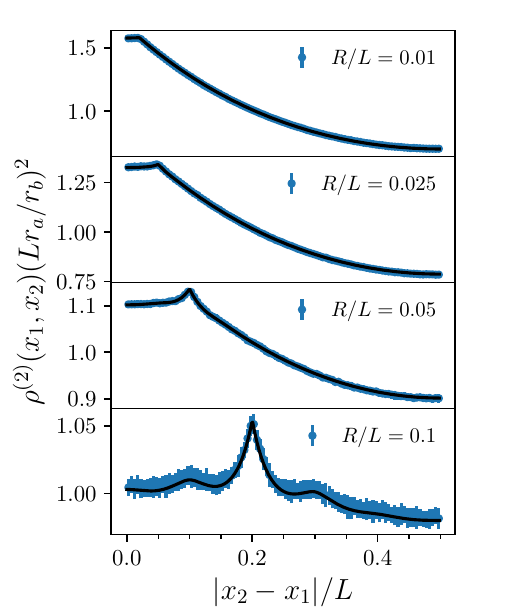}
  \caption{
    Two-body correlation function of the division, death, and diffusion (DDD) model in one dimension under periodic boundary conditions (period length $L$) with division separation $2R$.
    Markers are measured directly from simulation in LAMMPS \cite{LAMMPS,POP_package} (error bars are standard error of the mean).
    Solid lines are analytical results from the $2$-marginal probability distributions.
    Other parameters are fixed at $L=1$, $r_b/r_a = 100$, and $D/(L^2r_a)= 0.1$.} \label{fig:l2density}
\end{figure}

From Sec.~\ref{sec:1-marginal}, we know that the matrix $\bm{A}^{(1)}_{N\times N}(k_1+k_2)$ has a zero eigenvalue when $k_1+k_2=0$.
The condition $k_1+k_2=0$ is consistent with the translational symmetry of the system, 
which dictates that the probability distribution $p_n^{(2)}(x_1,x_2)$ depends solely on $x_2-x_1$ in the steady state.
The corresponding eigenvector for this zero eigenvalue is $\bm{\Pi}_N=\alpha(N)[\pi_1,\pi_2,\dots,\pi_N]^T$, where $\pi_n$ is given in Eq.~(\ref{eq:pi_n}).
$\alpha(N)$ is some normalization factor and in the limit $N\rightarrow\infty$, $\alpha(N)\rightarrow1$.
Thus, from the proposition above, it follows that zero is also an eigenvalue of the larger matrix $\bm{A}^{(1)+(2)}_{(2N-1)\times(2N-1)}(k_1,-k_1)$.
From numerical investigations, $\bm{A}^{(2)}_{(N-1)\times(N-1)}$ does not appear have a zero eigenvalue and is assumed to be invertible. 
Using the inverse, we can find the zero-eigenvector of the matrix $\bm{A}^{(1)+(2)}_{(2N-1)\times(2N-1)}(k_1,-k_1)$:
\begin{align}
&\bm{c}^{(1)+(2)}(k_1,k_2;t\rightarrow\infty)  \nonumber\\
&=\begin{bmatrix}
\bm{\Pi}_N  \\
-\left(\bm{A}^{(2)}_{(N-1)\times(N-1)}\right)^{-1}\bm{B}_{(N-1)\times N}\bm{\Pi}_N 
\end{bmatrix} \delta_{k_1,-k_2}.
\end{align}
The steady state $2$-marginal distribution $p_n^{(2)}(x_1,x_2;t\rightarrow\infty)$ can then be obtained by taking the inverse Fourier transform of the expression above.
This has to be done using numerical integration, however, for the special case of $R=0$, a closed-form expression for $p_n^{(2)}(x_1,x_2;t\rightarrow\infty)$ can be obtained:
\begin{equation}\label{eq:l2_p_n_R_is_0}
  p_n^{(2)}(x_1,x_2)
  = \frac{\pi_n}{2L^2}\frac{\omega}{\sinh(\frac{\omega}{2})}\cosh\left( \frac{\omega|x_1-x_2|}{L}-\frac{\omega}{2} \right),
\end{equation}
where $\omega = \sqrt{L^2 r_a/D}$.
For $R=0$, the $2$-marginal entropy production rate can also be computed and is found to be divergent, $\dot{S}_\text{irr}^{(2)}=\infty$ (see Appendix~\ref{app:2-marginal-Sdot}).
In contrast, the $1$-marginal entropy production rate vanishes for $R=0$, i.e., $\dot{S}_\text{irr}^{(1)}=0$ [see Eq.~(\ref{eq:1m_epr})].
This highlights that the rate of entropy production strongly depends on the number of degrees of freedom being tracked, as expected.

A more useful quantity is the $2$-body correlation function as defined in Eq.~(\ref{eq:lbodydensity}):
\begin{equation}
\rho^{(2)}(x_1,x_2) = \sum_{n=2}^\infty (n^2-n) p_n^{(2)}(x_1,x_2).
\end{equation}
Translation and inversion symmetries imply that $\rho^{(2)}(x_1,x_2)$ is solely a function of $|x_2-x_1|$.
Physically, $L^2\rho^{(2)}(|x_1-x_2|)/\left<n\right>_\text{ss}^2$ is the probability that we find another particle at a distance $|x_2-x_1|$ relative to a given particle at $x_1$.
{\color{black}Fig.~\ref{fig:l2density} shows the two-body correlation function as a function of the pairwise distance $|x_1-x_2|$.
Interestingly, the correlation function displays a peak at $|x_1-x_2|=2R$, which corresponds to the separation between two daughter particles immediately following a division event.
This peak indicates the most probable distance at which one is likely to find another particle relative to a given reference particle.}

\subsection{Brownian dynamics simulations}

To verify our results, we perform Brownian dynamics simulations of the DDD model, 
where the diffusing particles at positions $\{x\}_{n(t)}$ perform simple diffusion according to the stochastic differential equations
\begin{equation}
  \frac{dx_i}{dt} = \sqrt{2D}\xi_i(t), \quad i=1,\dots,n(t),
\end{equation}
where $n(t)$ is the total number of particles at time $t$.
$\xi_i(t)$ are white-noise with zero mean and variance $\langle\xi_i(t)\xi_j(t')\rangle=\delta_{ij}\delta(t-t')$.
The diffusion dynamics is coupled to division and death dynamics according to the rules described in Fig.~\ref{fig:DDD}.
{\color{black}
For all simulations, we initialize the system with a single particle at the origin and allow it to evolve towards a steady state.
We verify that the system has reached steady state by monitoring the number of particles 
$n(t)$ and confirming that it fluctuates around its steady state average value, 
$n(t)\sim\left<n\right>_\text{ss}$.
To compute the correlation function, we construct a histogram of the distances between all particle pairs.
This procedure is repeated over multiple simulations with independent noise realizations to obtain the ensemble average.
}
The simulations results, shown in Fig.~\ref{fig:l2density} as solid markers, demonstrate excellent agreement with the analytical results (solid lines) from the previous section.

\begin{figure}
  \def\svgwidth{0.5\textwidth}
\begingroup%
  \makeatletter%
  \providecommand\color[2][]{%
    \errmessage{(Inkscape) Color is used for the text in Inkscape, but the package 'color.sty' is not loaded}%
    \renewcommand\color[2][]{}%
  }%
  \providecommand\transparent[1]{%
    \errmessage{(Inkscape) Transparency is used (non-zero) for the text in Inkscape, but the package 'transparent.sty' is not loaded}%
    \renewcommand\transparent[1]{}%
  }%
  \providecommand\rotatebox[2]{#2}%
  \newcommand*\fsize{\dimexpr\f@size pt\relax}%
  \newcommand*\lineheight[1]{\fontsize{\fsize}{#1\fsize}\selectfont}%
  \ifx\svgwidth\undefined%
    \setlength{\unitlength}{243.77952756bp}%
    \ifx\svgscale\undefined%
      \relax%
    \else%
      \setlength{\unitlength}{\unitlength * \real{\svgscale}}%
    \fi%
  \else%
    \setlength{\unitlength}{\svgwidth}%
  \fi%
  \global\let\svgwidth\undefined%
  \global\let\svgscale\undefined%
  \makeatother%
  \begin{picture}(1,1.1627907)%
    \lineheight{1}%
    \setlength\tabcolsep{0pt}%
    \put(0,0){\includegraphics[width=\unitlength,page=1]{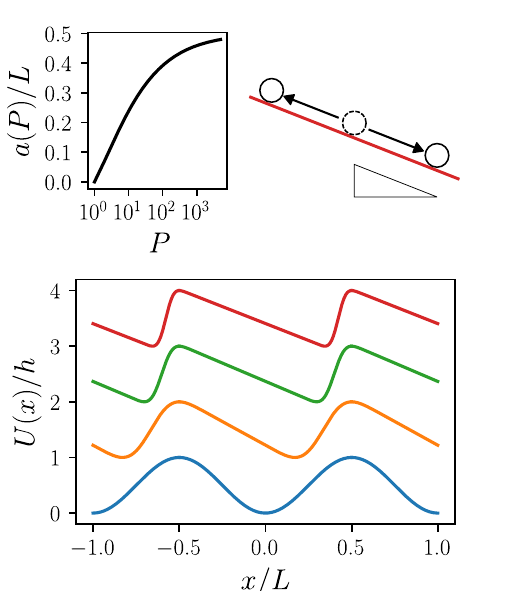}}%
    \put(0.76746025,0.81866818){\color[rgb]{0,0,0}\rotatebox{-21}{\makebox(0,0)[lt]{\lineheight{1.25}\smash{\begin{tabular}[t]{l}$R$\end{tabular}}}}}%
    \put(0.61274941,0.79481923){\color[rgb]{0,0,0}\makebox(0,0)[lt]{\lineheight{1.25}\smash{\begin{tabular}[t]{l}$\Gamma \sigma U'$\end{tabular}}}}%
    \put(0.76693255,0.75408261){\color[rgb]{0,0,0}\makebox(0,0)[lt]{\lineheight{1.25}\smash{\begin{tabular}[t]{l}$\sigma$\end{tabular}}}}%
    \put(0,0){\includegraphics[width=\unitlength,page=2]{projection.pdf}}%
    \put(0.60446872,0.88335823){\color[rgb]{0,0,0}\rotatebox{-21}{\makebox(0,0)[lt]{\lineheight{1.25}\smash{\begin{tabular}[t]{l}$R$\end{tabular}}}}}%
    \put(0,0){\includegraphics[width=\unitlength,page=3]{projection.pdf}}%
    \put(0.19767442,1.06976744){\color[rgb]{0,0,0}\makebox(0,0)[lt]{\lineheight{1.25}\smash{\begin{tabular}[t]{l}(a)\end{tabular}}}}%
    \put(0.19767442,0.58139535){\color[rgb]{0,0,0}\makebox(0,0)[lt]{\lineheight{1.25}\smash{\begin{tabular}[t]{l}(b)\end{tabular}}}}%
  \end{picture}%
\endgroup%

  \vspace{0pt}
  \caption{Ratchet potential applied to the DDD process.
    In (b), the applied, $L$-periodic ratchet potential, $U(x)$ in Eq.~(\ref{eq:ratchetpotential}),
    is shown for integer values $P$ which dictate the location of the minima in $x$.
    From bottom to top, $P=1$ (blue), $P=5$ (orange), $P=20$ (green), and $P=40$ (red). 
    A constant offset is added to better visualise the different curves shown. 
    The location of the minimum in $x\in[0,L/2]$, $a(P)$ in Eq.~(\ref{eq:asym_P}), is shown in (a). 
    The inset of (b) illustrates the geometric projection of a division event onto the local slope of the potential $U(x)$ with $\sigma=R/\sqrt{1+\left[\Gamma U'(x)\right]^2}$ as the projected separation radius, see Eq.~(\ref{eq:local_split}).}
  \label{fig:ratchet}
\end{figure}

\section{A minimal model of particle current via division, death and diffusion \label{sec:geoproj}}

As a final application of our formalism, we examine the DDD model within an asymmetric ratchet potential $U(x)$, as illustrated in Fig.~\ref{fig:ratchet}.
This setup serves as a minimal model of non-motile cells confined in a periodic, asymmetric channel, where they undergo division, death and diffusion. 
Previous simulations of dividing particles in a two-dimensional ratcheted channel have demonstrated the emergence of a non-zero macroscopic particle current, driven by steric interactions among the particles and with the channel walls \cite{D1SM00928A}. 
To capture the essential physics of this process, we consider a one-dimensional system of non-interacting particles subject to a periodic ratchet potential $U(x)$~\cite{Burada2009Diffusion}.
This is achieved by incorporating a drift term into the current operator $\hat{j}_{x_i}$ in Eq.~(\ref{eq:hetero-master}-\ref{eq:j-f}):
\begin{equation}
\hat{j}_{x_i} = -D\partial_{x_i} + \gamma U'(x_i), \label{eq:current_operator}
\end{equation}
where $D>0$ is the diffusion constant and $\gamma>0$ is the friction coefficient.
If fluctuation-dissipation holds, we have the relationship $D=k_BT/\gamma$, where $T$ represents the (effective) temperature of the surrounding heat bath, however, this is not always true in an active system.
Additionally, we extend our previous DDD model to account for a geometric projection of particle division along the slope of the potential $U(x)$, 
see Fig.~\ref{fig:ratchet}(b).
This modification prevents particles from `tunneling' through the potential barrier during division. 
Consequently, when a particle divides at position $x$, the two daughter particles are positioned at
\begin{equation}\label{eq:local_split}
  f_{\pm R}(x) = x \pm \frac{R}{\sqrt{1+\left[\Gamma U'(x)\right]^2}}
\end{equation}
where $R$ is now the maximum splitting radius 
(in contrast to the Sec.~\ref{sec:DDD} where particles were always shifted by $\pm R$ relative to the parent particle).
The phenomenological parameter $\Gamma$ couples the projected splitting radius $\sigma=R/\sqrt{1+\left[\Gamma U'(x)\right]^2}$ to the ratchet potential $U(x)$. 
The splitting radius is given by $R$ only where the force due to the potential is zero $U'(x)=0$ or if $\Gamma=0$. 
Otherwise, the model is given by Eq.~(\ref{eq:hetero-master}) with the new definition of $f_{\pm R}(x)$ given in
Eq.~(\ref{eq:local_split}) and the additional drift term in the current operator in Eq.~(\ref{eq:current_operator}).

Our aim is to determine the average density $\rho(x,t)$, which can be obtained from the set of $1$-marginal probability distributions $\rho(x,t)=\sum_{n=1}^\infty \, n p_n^{(1)}(x,t)$, see Eq.~(\ref{eq:lbodydensity}).
The constraints of Eqs.~(\ref{eq:widetilde_constraint}-\ref{eq:leftarrow_constraint}) are still satisfied in spite of the addition of a non-zero external field and the new splitting rule in Eq.~(\ref{eq:local_split}).
Therefore, one may multiply Eq.~(\ref{eq:formalmarkov_p_n_l_dynamics}) by $n$ and sum over all integers $n\ge1$ to determine $\rho(x,t)$. 
The resulting dynamics of $\rho(x,t)$ depends explicitly on the $1$-marginal probability distributions $p_n^{(1)}(x,t)$. 
To close the equation for $\rho(x,t)$, we make the approximation $\langle n\rangle/\langle n^2\rangle\approx \sum_m m p_m^{(1)}(x)/\sum_n n^2p_n^{(1)}(x)$, i.e., the ratio of average particle number to its square scales the same locally as it does globally (in steady-state).
This approximation is explained Appendix~\ref{app:geom_proj} and justified in Appendix~\ref{app:ratchet_potential} for a smooth ratchet potential.
The steady-state average density is then given by the differential equation
\begin{equation}\label{eq:rho_full_sink}
  \partial_x J[\rho(x),U(x)] = r_bS[\rho(x),U(x)],
\end{equation}
where the divergence of the current
\begin{equation}
  J[\rho,U]= -\frac{1}{\gamma} U'(x)\rho(x) - D \rho'(x)
\end{equation}
balances a source/sink term
\begin{equation}\label{eq:full_sink}
    S[\rho,U] \equiv \left[ \sum_{\sigma\in\{-R,R\}} \int dy \, \delta(f_{\sigma}(y)-x)\rho(y) \right] - 2\rho(x)
\end{equation}
driven by division and death. 
Note that a density equation similar to Eqs.~(\ref{eq:rho_full_sink}-\ref{eq:full_sink}) is usually the starting point of dynamic density functional theory~\cite{teVrugt02042020}.
In this paper, we derive the time evolution of the density (and two-body correlation function) from the full Fokker-Planck equation through a hierarchy of marginal probability distributions.
As we show in the Appendix~\ref{app:geom_proj} and \ref{app:ratchet_potential},
if one requires that the maximum splitting radius is much smaller than the channel period,  $R/L\ll 1$, 
the differential equation may be recast as
\begin{equation}\label{eq:ode_rho}
  0 = \partial_x\left\{\left[\frac{1}{\gamma}\partial_x \tilde{U}(x) + \tilde{D}(x)\partial_x\right]\rho(x)\right\} + O\left(\frac{R^3}{L^3}\right)
\end{equation}
with the definitions of an effective potential and diffusion coefficient,
 \begin{align}
    \tilde{U}(x) &\equiv U(x) +  \frac{\gamma r_b R^2}{1 + \left[\Gamma U'(x)\right]^2}, \label{eq:effective-U}\\
    \tilde{D}(x) &\equiv D + \frac{r_bR^2}{1 + \left[\Gamma U'(x)\right]^2}, \label{eq:effective-D}
 \end{align}
respectively. The solution of Eq.~(\ref{eq:ode_rho}) can be solved using quadrature~\cite{risken2012fokker} with the normalisation requirement
\begin{equation}
  \langle n \rangle = \int_{-L/2}^{L/2} dx \, \rho(x) 
\end{equation}
and periodic boundary conditions $\rho(x+L)=\rho(x)$.

\begin{figure}
  \def\prefix {../experiments/2024_03_17/more_terms/}
  \includegraphics[width=0.4\textwidth]{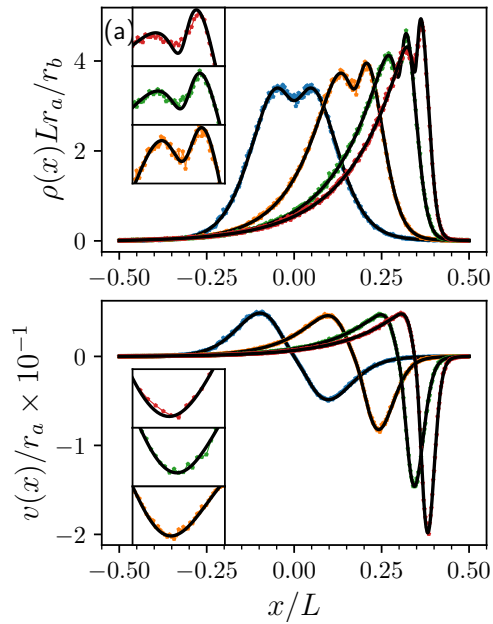}
  \caption{(a) average density $\rho(x)$ and (b) average local velocity (per particle per length) $v(x)$ of the DDD process subject to an external field and geometric projection of particle division in steady state. 
Our analytical results are represented by black solid lines, while direct simulations performed using LAMMPS~\cite{LAMMPS, POP_package} are shown as solid markers.
    From left to right in terms of the maxima of the curves, the externally applied ratchet potential is asymmetrically skewed such that its minimum is at 
    $x/L=0$ (blue), $x/L = 0.17$ (orange), $x/L = 0.30$ (green) and $x/L = 0.34$ (red). 
    Other parameters are fixed at  $R/L=0.015$, $r_b/r_a = 400$, $\Gamma D\gamma/L=0.15$,
    $D/(L^2 r_a)=0.15$,  and $D\gamma/h = 0.15$. 
    Insets highlight differences between the analytics and simulation at their most prominent regions of disagreement.} \label{fig:asym_prob}
\end{figure}

As mentioned in the previous paragraph, Eq.~(\ref{eq:ode_rho}) assumes that average particle number and its square scale in a spatially invariant way,
and such assumption is not obvious a-priori. 
In Fig.~\ref{fig:asym_prob}, we compute $\rho(x)$ using Eq.~(\ref{eq:ode_rho}) and compare it to the average (steady-state) density measured in Brownian
Dynamics simulations, where particle $x_i(t)$ obeys the stochastic differential equation
\begin{equation}\label{eq:geo_langevin}
  \frac{dx_i}{dt} = -\frac{1}{\gamma}U'(x_i) +  \sqrt{2D}\xi_i(t), \quad i=1,\dots,n(t),
\end{equation}
where $n(t)$ is the number of particles, which changes due to division and death events.
Each division event shifts the daughter particles by a separation radius, which is geometrically projected onto the local slope of $U(x)$, according to Eq.~(\ref{eq:local_split}). 
For the potential $U(x)$ itself, we choose a smooth ratchet potential
\begin{equation} \label{eq:ratchetpotential}
  U(x) = \frac{h}{2}\bigg(\frac{\sum_{k=1}^P\binom{2P}{P-k}\sin(2\pi k x/L)/k}{\sum_{k=1}^P\binom{2P}{P-k}\sin(k \pi (1-a(P)/L))/k }+1\bigg),
\end{equation}
which is differentiable everywhere and periodic with period $L$.
$h$ is  the energy scale of the potential and $\binom{\cdots}{\cdots}$ is the binomial coefficient. 
The minimum of this potential is located at
\begin{equation}\label{eq:asym_P}
  a(P) = \frac{L}{2}-\frac{2L}{\pi}\cos^{-1}\bigg(\frac{1}{2}\binom{2P}{P}^{\frac{1}{2P}}\bigg).
\end{equation}
which is monotonically increasing in $P$ (see Figure \ref{fig:ratchet}).
The integer $P\in\mathbb{Z}^{+}$ in Eq.~(\ref{eq:ratchetpotential}) controls the asymmetry of the ratchet with 
$a(P=1)=0$ (completely symmetric) and $a(P\to\infty)\to L$. 
In Fig.~\ref{fig:asym_prob}(a) we find good agreement between the calculated marginal density and Brownian Dynamics simulations. 
Similarly, in Fig.~\ref{fig:asym_prob}(b) our calculation of the steady-state average local velocity per particle per length,
$v(x)\equiv -\gamma^{-1} \rho(x) U'(x)/\langle n \rangle $, is in agreement with simulation.

\begin{figure}
  \def\prefix{../experiments/2024_03_19/ratchet_entropyproduction/}
  \includegraphics[width=0.5\textwidth]{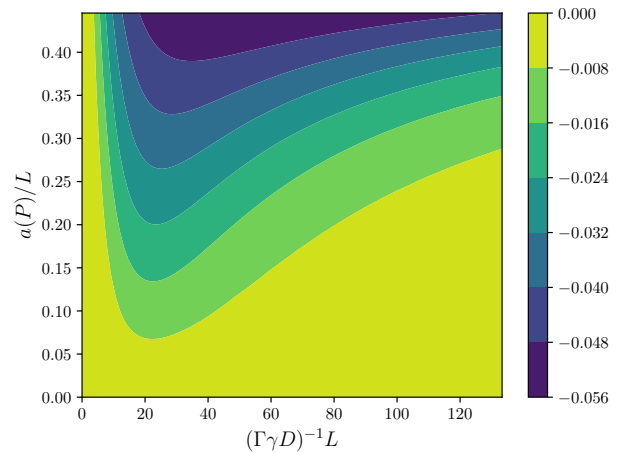}
  \caption{Global velocity per particle (in units of $L r_a$)
    for the geometrically projected DDD process. The ratchet potential asymmetry $a(P)/L$ is varied
    by changing the integer $P\in[1,40]$, and inverse dimensionless geometrical coupling
    parameter $(\Gamma D\gamma)^{-1}L$ is changed while holding all other parameters
    fixed at  $R/L=0.015$, $r_b/r_a=400$, $D/(L^2 r_a)=0.15$,
    and $D\gamma/h = 0.15$.} \label{fig:current}
\end{figure}

Given the excellent agreement between the calculated $\rho(x)$ and $v(x)$ with simulation, 
we compute the average global velocity of this geometrically projected DDD process by integrating $v(x)$ over the period $L$, 
enabling us to scan parameter space much more efficiently than using direct simulation. 
In Fig.~\ref{fig:current} we show that one does indeed get a macroscopic particle current with our minimal model of cells in a ratchet \cite{D1SM00928A} due to the broken symmetry of the geometrically projected splitting radius.
Of course, to quantitatively match our model to previous simulations and/or experiments, one would need to include, e.g., particle interactions, two-dimensional correlations due to division, and explicit wall interactions, which are beyond the scope of this paper.

\section{Discussion and conclusions.}

In this work, we have presented a formalism which suitably describes systems with a state space that has both discrete and continuous degrees of freedom. 
In the context of existing literature, textbooks allude to the possibility of extending formalism of, e.g., the Fokker-Planck equation or the master equation to hybrid state space, but in practice this is not done explicitly~\cite{gardiner1985handbook,van2011stochastic,risken2012fokker}.
Additionally, in this manuscript we specify that the number of continuous degrees of freedom is itself the discrete degree of freedom, i.e., having two continuous degrees of freedom $x_1$ and $x_2$ implies the state $\{2,x_1,x_2\}$.

We have studied a toy model of division, death, and diffusion (DDD) to illustrate a simple application of our formalism. 
In the absence of an external potential, the set of probability distribution functions $\{p_n(\{x\}_n,t)\}$ can be solved recursively using a hierarchy similar to the BBGKY hierarchy of deterministic (Hamiltonian) mechanics which we solve analytically up to second order (two-body probability distributions). 
The resulting two-body correlation function demonstrates excellent agreement with numerical results obtained from direct Brownian dynamics simulations.

Additionally, we also computed the $1$-marginal and the $2$-marginal entropy production rate,
revealing a strong dependence on the number of degrees of freedom being tracked.
For example when $R=0$, the $1$-marginal entropy production rate vanishes whereas the $2$-marginal entropy production rate is infinite.
The entropy production rate is also sensitive to the nature of the division separation—whether it is random or fixed at $\pm R$ relative to the original position.
We demonstrate this by treating the delta function as the singular limit of a Gaussian distribution in our calculation of the one-marginal entropy production rate.

In the presence of an external potential $U(x)$, we derive the effective equation of state for the average particle density $\rho(x)$. 
In the minimal model, an asymmetric ratchet potential combined with geometric projection is sufficient to generate a macroscopic current. 
Despite its simplicity, this coupling produces a non-zero particle flow in qualitative agreement with particle-based simulations.

Future extensions of the model, or applications of the formalism presented here—particularly those motivated by experiments on confluent cell layers—will almost certainly require incorporating particle interactions. 
This would necessitate a closure relation analogous to the molecular chaos hypothesis in kinetic theory to enable further analytical progress. 
Another promising extension involves introducing an absorbing state transition into the vacuum state ($n=0$), enabling the study of directed percolation and related models~\cite{PhysRevLett.133.067103,Hinrichsen01112000}.

Overall, the framework developed in this work provides a theoretical foundation for constructing more complex models that integrate division, death, and diffusion, bridging microscopic dynamics with hydrodynamic descriptions in terms of reaction-diffusion systems.

\section{Acknowledgements}
We acknowledge support from EPSRC grant no. EP/W027194/1.
ET would like to thank Daniel Sussman, Rob Jack, and Mike Cates for useful discussions.
ET would like to thank the Isaac Newton Institute for Mathematical Sciences, Cambridge, for support and hospitality during the SPL programme, 
funded by EPSRC grant no EP/R014604/1.

\appendix

\section{Entropy production in a hybrid state system \label{app:entropy}}

In this section, we shall use a short-hand notation to simplify the presentation:
\begin{equation} 
  \begin{split}
     p(n_j) &= p_{n_j}(\{x(t_j)\}_{n_j},t_j) \\
     G(n_j \rightarrow n_k) &= G_{n_jn_k}(\{x(t_j)\}_{n_j}\rightarrow\{x(t_k)\}_{n_k};\tau) \\
     \int dn_j &= \sum_{n_j=1}^\infty \int d\{x(t_j)\}_{n_j}
  \end{split}
\end{equation}
Substituting Eqs.~(\ref{eq:path}) and (\ref{eq:reverse_def}) into (\ref{eq:entropy_change}), we get: 
\begin{widetext} 
\begin{align} \label{eq:entropy_change_1}
     \Delta{\mathcal{S}}_M &= \underbrace{\left( \int dn_0 \int dn_1 \dots \int dn_M \right)}_{\sum_{\bm{X}_M}}
     \underbrace{ p(n_0)\prod_{k=0}^{M-1} G(n_k\rightarrow n_{k+1}) }_{\mathcal{P}(\bm{X}_M;t_0,\Delta t)}
     \ln\left( \frac{p(n_0)\prod_{j=0}^{M-1} G(n_j\rightarrow n_{j+1}) }
     		       {p(n_M)\prod_{j=0}^{M-1} G(n_{j+1}\rightarrow n_j) } \right)  \\
      &= \sum_{j=0}^{M-1}  \left( \int dn_0 \int dn_1 \dots \int dn_M \right)
     p(n_0)\prod_{k=0}^{M-1} G(n_k\rightarrow n_{k+1}) 
      \ln\left( \frac{p(n_j) G(n_j\rightarrow n_{j+1}) }
     		       {p(n_{j+1}) G(n_{j+1}\rightarrow n_j) } \right) \\
      &= \sum_{j=0}^{M-1} \int dn_0 \int dn_j \int dn_{j+1}  \int dn_M \, p(n_0)G(n_0\rightarrow n_j)G(n_j\rightarrow n_{j+1})G(n_{j+1}\rightarrow n_M)
           \ln\left( \frac{p(n_j) G(n_j\rightarrow n_{j+1}) }{ p(n_{j+1}) G(n_{j+1}\rightarrow n_j) } \right),
\end{align}
\end{widetext}
where we have used the Chapman-Kolmogorov relation to get the last line in the equation above:
\begin{equation}
    G(n_0\rightarrow n_j) = \left( \int dn_1\dots\int dn_{j-1} \right)  \prod_{k=0}^{j-1} G(n_k\rightarrow n_{k+1}).
\end{equation}
Finally we use the Markov property on the last line of Eq.~(\ref{eq:entropy_change_1}):
\begin{align}
    \int dn_0 \, p(n_0)G(n_0\rightarrow n_j) &= p(n_j) \\
    \int dn_j \, G(n_0\rightarrow n_j) &= 1
\end{align}
so that Eq.~(\ref{eq:entropy_change_1}) simplifies into:
\begin{align}
\Delta{\mathcal{S}}_M &= \sum_{j=0}^{M-1} \int dn_j \int dn_{j+1} \, p(n_j)G(n_j\rightarrow n_{j+1}) \nonumber\\
     &\times\ln\left( \frac{p(n_j) G(n_j\rightarrow n_{j+1}) }{ p(n_{j+1}) G(n_{j+1}\rightarrow n_j) } \right),
\end{align}
Finally, we take the limit $M\rightarrow\infty$ and $\tau\rightarrow0$, the summation $\sum_{j=0}^{M-1}\tau$ then becomes a Riemann integral $\int_{t_0}^{t_0+\Delta t}dt$.
We therefore identify the integrand to be the rate of entropy production~\cite{Gaspard2004,e22111252} at time $t_j = t_0 + j\tau$:
\begin{align}
    &\dot{S}_\text{irr}(t_j) = \lim_{\tau\rightarrow\infty}\frac{1}{\tau} \sum_{n_j=1}^\infty \sum_{n_{j+1}=1}^\infty \int d\{x\}_{n_j}\int d\{y\}_{n_{j+1}} \nonumber\\
    &\times p_{n_j}(\{x\}_{n_j},t_j) G_{n_jn_{j+1}}(\{x\}_{n_j}\rightarrow\{y\}_{n_{j+1}};\tau) \nonumber\\
    &\times \ln\left( \frac{p_{n_j}(\{x\}_{n_j},t_j) G_{n_jn_{j+1}}(\{x\}_{n_j}\rightarrow\{y\}_{n_{j+1}};\tau)}
    				   {p_{n_{j+1}}(\{y\}_{n_{j+1}},t_{j+1}) G_{n_{j+1}n_j}(\{y\}_{n_{j+1}}\rightarrow\{x\}_{n_j};\tau)} \right)
\end{align}
After relabelling $t_j\rightarrow t$, $t_{j+1}\rightarrow t$, $\{x\}_{n_j}\rightarrow\{x\}_m$ and $\{y\}_{n_{j+1}}\rightarrow\{y\}_n$, we finally get Eq.~(\ref{eq:entropy_production}) in the main text.

\section{$2$-marginal entropy production rate \label{app:2-marginal-Sdot}}

In the case of $R=0$, the steady state $2$-marginal probability distributions have the simple analytical form of Eq.~(\ref{eq:l2_p_n_R_is_0}).
We now show that the $2$-marginal entropy production rate is infinite even when $R=0$. 
Just as in the $1$-marginal case, only the off-diagonal terms in the summation of the $2$-marginal entropy production [Eq.~(\ref{eq:S-marginal})] are non-zero. 
Therefore, to show that $\dot{S}_i^{(2)}=\infty$ for $R=0$, one just needs to compute  the first off-diagonal contribution to the marginal entropy production rate
($1\leftrightarrow 2$ particles) and prove that it is infinite. 
Using Eq.~(\ref{eq:l2_p_n_R_is_0}), this first term is
\begin{widetext}
  \begin{align}\label{eq:2m_12epr}
      &\lim_{\tau\to0}\frac{1}{\tau}\int dy_1 dx_1dx_2 \,
      G(\{y\}_1\to\{x\}_2;\tau)p_1(\{y\}_1)
      G(\{x\}_2\to\{y\}_1;\tau)p_2(\{x\}_2)
      \ln\left[\frac{G(\{y\}_1\to\{x\}_2;\tau)p_1(\{y\}_1)}{G(\{x\}_2\to\{y\}_1;\tau)p_2(\{x\}_2)}\right]  \nonumber\\
      &=\int dy_1 dx_1dx_2 \,
      \ln\left[\frac{4 L \sinh(\omega/2)\delta(x_1-y_1)\delta(x_2-y_1)}{\omega\cosh(\omega|x_2-x_1|/L-\omega/2)\left[\delta(x_2-y_1)+\delta(x_1-y_1)\right]}\right] \nonumber\\
      & \times \left\{\frac{r_b\pi_1}{L}\delta(x_1-y_1)\delta(x_2-y_1) 
      - \frac{r_a\pi_2}{L^2}\frac{\omega}{\sinh(\omega/2)}\cosh\left(\frac{\omega|x_2-x_1|}{L}-\frac{\omega}{2}\right)
        \left[\delta(x_1-y_1) + \delta(x_2-y_1)\right]\right\},
  \end{align}
\end{widetext}
where $\omega=\sqrt{L^2r_a/D}$.
As done for the $1$-marginal case, if one approximates the delta-functions in Eq.~(\ref{eq:2m_12epr}) as infinitesimal width Gaussians, 
we find that the $2$-marginal entropy production rate is infinite even when $R=0$. 
Notably, if one takes the infinite diffusion limit $D\to\infty$, 
the entropy production rate remains infinite even though the $2$-marginal distribution $p_n^{(2)}(x_1,x_2)$ becomes spatially homogeneous. 
Therefore, inclusion of additional information on spatial locations of particles in the DDD process
changes the entropy production rate even when the system is spatially homogeneous,
i.e., the entropy production is not simply given by Eq.~(\ref{eq:homo_entropy}).

\section{DDD geometrically coupled to $U(x)$ \label{app:geom_proj}}

In this Appendix, we derive Eq.~(\ref{eq:ode_rho}) with the effective potential and diffusion coefficients given in Eqs.~(\ref{eq:effective-U}-\ref{eq:effective-D}). 
We start with the DDD dynamics defined in Eq.~(\ref{eq:hetero-master}) but with the general current operator $\hat{j}_{x_i}$ given in Eq.~(\ref{eq:current_operator}) and general splitting rule $f_{\pm R}(x)$ given in Eq.~(\ref{eq:local_split}). 
Integrating Eq.~(\ref{eq:hetero-master}) over all but $x_1$, we find that the dynamics of $1$-marginal distribution functions obey the set of equations
\begin{align}\label{appeq:p_n1_full}
    &\partial_t p_n^{(1)}(x,t) + \partial_x\left[\hat{j}_xp_n^{(1)}(x,t) \right] \nonumber\\
&=r_a(\mathbb{E}-1)n(n-1)p_n^{(1)}(x,t) +(\mathbb{E}^{-1}\!\!-1)np_n^{(1)}(x,t) \nonumber\\
  &+\frac{r_b(n-1)}{n}\int dy\bigg[ \delta(f_{+R}(y)-x) +\delta(f_{-R}(y)-x) \nonumber\\
    & \quad\quad\quad\quad\quad\quad   - 2\delta(y-x)\bigg] p_{n-1}^{(1)}(y,t).
\end{align}
where we have defined the step operator $\mathbb{E}[f(n)]=f(n+1)$ and its inverse $\mathbb{E}^{-1}[f(n)] = f(n-1)$.
The terms which shift the particle by Eq.~(\ref{eq:local_split}) can be integrated,
\begin{align}
    &\int dy \, \delta(f_{\pm R}(y)-x)p_{n-1}^{(1)}(y,t)  \nonumber\\
   =&\sum_{\alpha=1}^S\frac{p_{n-1}^{(1)}(f_{\pm R,\alpha}^{-1}(x),t)}{1\pm R Q'(f_{\pm R,\alpha}^{-1}(x))} \nonumber\\
   =& \sum_{\beta=0}^{\infty}\frac{(\mp R)^{\beta}}{\beta !}(\partial_x)^{\beta}\big(p_{n-1}^{(1)}(x,t)Q^{\beta}(x)\big) \label{appeq:multiple_options}  
\end{align}
where we have defined
\begin{equation}\label{appeq:Q}
  Q(x) = \sqrt{\frac{1}{1+\left[\Gamma U'(x)\right]^2}}
\end{equation}
for convenience, c.f., Eq.~(\ref{eq:local_split}).
The second line in Eq.~\ref{appeq:multiple_options}  is obtained by a substitution of variables to $v=f_{\pm R}(y)$ and then direct integration over $v$. $f_{\pm R,\alpha}^{-1}(x)$ are the $S$ inverses of Eq.~(\ref{eq:local_split}) over the period $x\in[-L/2,L/2]$, i.e.,
$f_{\pm R,\alpha}^{-1}(x) \in \{ y\: :\: x = f_{\pm R}(y), \: x\in[-L/2,L/2]\}$.
(Note that $f_{\pm R,\alpha}^{-1}(x)$ is, in general, a multi-valued function.)
The last line in Eq.~(\ref{appeq:multiple_options}) follows from formally Taylor expanding $\delta(f_{\pm R}(y)-x)$ about small $R$. 
Both forms are equivalent. 
We will use the direct substitution form to motivate what parameter regimes can be reasonably approximated by the second order differential equation (\ref{eq:ode_rho}) in the main text. 
The Taylor expansion form will be used otherwise as it is simpler to use in algebraic manipulations. 
In particular, it highlights that when $R=0$ and fluctuation-dissipation theorem $D=k_BT/\gamma$ holds, the steady-state solution for the $1$-marginal is $p_n^{(1)}(x)=\pi_n e^{-U(x)/k_BT}/Z$ where $Z$ is a normalisation factor, i.e., when the splitting radius is zero the population dynamics decouple from the spatial dynamics in steady state.

Multiplying Eq.~(\ref{appeq:p_n1_full}) by $n$ and then summing over all $n$, the average density obeys the differential equation
\begin{align}\label{appeq:rho1_full}
 &\partial_t \rho(x,t) + \partial_x \left[\hat{j}_x\rho(x,t)\right] \nonumber\\
    =& (r_a + r_b)\rho(x,t) - r_a\sum_{n=1}^{\infty} n^2p_n^{(1)}(x,t)-2r_b\rho(x,t) \nonumber\\
    &+ r_b \sum_{\sigma\in\{-R,R\}} \int dy \, \delta(f_{\sigma}(y)-x)\rho(y,t).
\end{align}
Therefore, $\rho(x,t)$ depends explicitly on the second moment of particle number. 
One either needs to solve for $p_n^{(1)}(x,t)$ first and then sum over $n$ to determine $\rho(x,t)$, or introduce an approximation to close Eq.~(\ref{appeq:rho1_full}).
We opt for the latter approach, taking
\begin{equation}\label{appeq:closure}
  \sum_{n=1}^{\infty} n^2 p_n^{(1)}(x,t)\approx \frac{\langle n^2(t)\rangle}{\langle n(t) \rangle} \rho(x,t) + O(R^2)
\end{equation}
to close Eq.~(\ref{appeq:rho1_full}), where $\langle n(t) \rangle$ and $\langle n^2(t) \rangle$ are the position independent averages of particle number and its square at time $t$.
Eq.~(\ref{appeq:closure}) is equivalent to the assumption that $p_n^{(1)}(x,t)$ is separable in
$n$ and $x$ to first order in $R^2$. 
The error introduced by this approximation is multiplied by $r_a$ in Eq.~(\ref{appeq:rho1_full}), 
so we expect our results to be valid so long as $r_b > r_a$ such that the final term in Eq.~(\ref{appeq:rho1_full}) still dominates the $O(R^2)$ correction. Since we choose $r_b\gg r_a$ in this manuscript, we expect this approximation to introduce negligible error in our calculations of $\rho(x)$. 
We will directly verify in Appendix \ref{app:ratchet_potential} that this approximation is reasonable.
Finally, by using Eq.~(\ref{appeq:closure}), cancellation occurs between the two terms in the second line of Eq.~(\ref{appeq:rho1_full}).
The closed equation for $\rho(x,t)$ is
\begin{align}\label{appeq:rho1_closed}
 &\partial_t \rho(x,t) + \partial_x\left[\hat{j}_x\rho(x,t)\right]  \nonumber \\
 = &r_b\bigg(
    \sum_{\alpha=1}^S\frac{\rho(f_{+R,\alpha}^{-1}(x),t)}{1+R Q'(f_{+R,\alpha}^{-1}(x))}  \nonumber\\
    & + \sum_{\alpha=1}^S\frac{\rho(f_{-R,\alpha}^{-1}(x),t)}{1-R Q'(f_{-R,\alpha}^{-1}(x))} - 2\rho(x,t) \bigg).
\end{align}
The terms on the right-hand side of this equation represent a source and a sink term due to division and death, given by $S[\rho,U]$ in Eq.~(\ref{eq:full_sink}). 
By expanding
this source/sink term to $O(R^2)$ using the second Taylor expansion approximation of the
splitting events, c.f., Eq.~(\ref{appeq:multiple_options}), we recover Eq.~(\ref{eq:ode_rho}) with the effective potential and diffusion coefficients given in
Eqs.~(\ref{eq:effective-U}-\ref{eq:effective-D}).

\begin{figure}
  \def\prefix{../experiments/2024_11_18/closure_assumption/}
  \includegraphics[width=0.5\textwidth]{{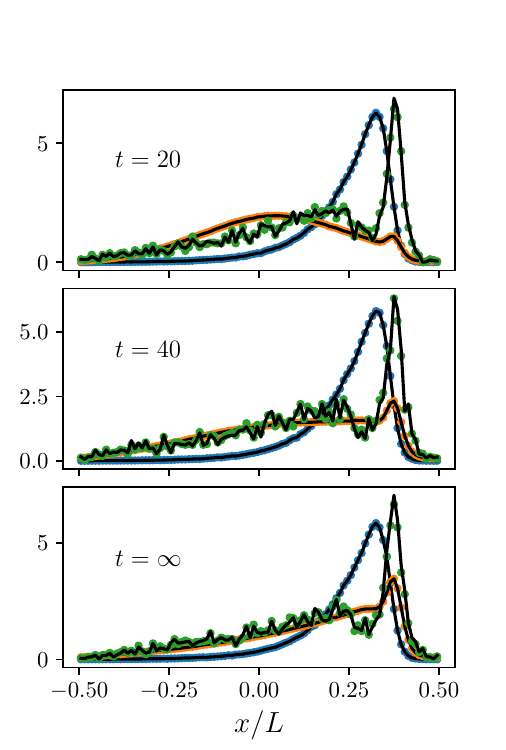}}
  \caption{Comparison of $\langle n^2\rangle\rho(x,t)/\langle n\rangle^3$ (dots)
    and $\sum_n n^2 p_n^{(1)}(x,t)/\langle n\rangle^2$ (black lines).
    Both quantities are measured directly from molecular dynamics simulations coupled to division and death events as discussed in the main text. 
    Different colours correspond to different values of the parameter 3-tuple $(r_a,r_b,L)$: 
    $(\num{e-3},0.4,10)$ (blue), $(\num{e-3},4,66)$ (orange), and $(10,40,10)$ (green). 
    Other parameters are set at $P=30$ (ratchet potential asymmetry corresponding to $a(P)/L\approx0.3$),
    $D=1$, $R=1$, $\Gamma = 2$, $h=7.9$, and $\gamma=1$.
  }\label{appfig:closure}
\end{figure}

\section{Smooth ratchet potential \label{app:ratchet_potential}}

In Appendix~\ref{app:geom_proj}, we introduced two approximations which allowed us to derive Eq.~(\ref{eq:ode_rho}) for general potential $U(x)$. 
Here, we justify these approximations specifically for the ratchet potential used in the main text [Eq.~(\ref{eq:ratchetpotential})].

The first approximation made is to assume Eq.~(\ref{appeq:closure}) is valid. 
We check in Fig.~\ref{appfig:closure} that Eq.~(\ref{appeq:closure}) is reasonable by directly simulating particle trajectories via Eq.~(\ref{eq:geo_langevin})(coupled to the geometrically projected division and death), and measuring $\rho(x,t)$, $\sum_n n^2p_n^{(1)}(x,t)$, $\langle n(t)\rangle$, and $\langle n^2(t)\rangle$.
We find excellent overlap between the left hand side (black lines) and right hand side (dots) of Eq.~(\ref{appeq:closure}). 
As discussed in Appendix~\ref{app:geom_proj},
when $r_a$ approaches $r_b$ and both are large, Eq.~(\ref{appeq:closure}) should begin to fail. 
We find good agreement when we take $r_a=10$ and $r_b=40$ (green dots and overlapping black line in Fig.~\ref{appfig:closure}). 
Therefore, Eq.~(\ref{appeq:closure}) is reasonable for all parameter regimes studied in this paper when $U(x)$ is given by Eq.~(\ref{eq:ratchetpotential}).

The second assumption made in deriving Eq.~(\ref{eq:ode_rho}) is that $R/L\ll 1$ such that the Taylor series expansion in Eq.~(\ref{appeq:multiple_options}) is sufficiently approximated by the first correction $O(R^2)$, or equivalently that there is a unique inverse of $f_{\pm R}(y)$ 
[so $S=1$ in Eq.~(\ref{appeq:multiple_options})]
and the summations in Eq.~(\ref{appeq:p_n1_full}) reduces to a single term. 
This depends strongly on the choice of potential $U(x)$ being used. 
We verify here the regions of parameter space which affect the Taylor expansion validity and unique invertibility of $f_{\pm R}(y)$
for the ratchet potential [Eq.~(\ref{eq:ratchetpotential})].
Since $f_{\pm R}(y)$ depends only on the dimensionless parameters $R/L$, $\Gamma h/L$, and the asymmetry number $P$ (see Fig.~\ref{fig:ratchet}a), these three parameters completely dictate the regions of validity for Eq.~(\ref{eq:ode_rho}).

\begin{figure}
  \def\prefix{../experiments/2024_11_25/finversion/}
  \includegraphics[width=0.5\textwidth]{{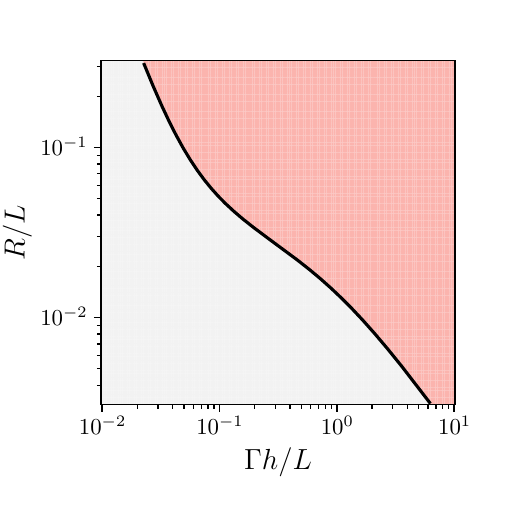}}
  \caption{Region in which Eq.~(\ref{eq:ode_rho}) is valid (grey shade) and when higher order terms in $R$ must be considered (red shade) for the asymmetry parameter $P=100$. 
  The black solid line is the criterion $R=1/\mathrm{max}(Q')$. 
  The three parameters $R/L$, $\Gamma h/L$, and $P$ determine whether the expansion of Eq.~(\ref{appeq:rho1_closed}) to second order in splitting radius $R$ is sufficient.
  }\label{appfig:parameter_region}
\end{figure}

The inverse of $f_{\pm R}(y)$ is unique if it is one-to-one, i.e., there is no maximum. 
In Fig.~\ref{appfig:parameter_region}, we show a map of the region where at least one maximum is present (absent) in light red (grey) when the ratchet asymmetry parameter $P=100$.
Therefore, the grey regions of Fig.~\ref{appfig:parameter_region} should permit one to use Eq.~(\ref{eq:ode_rho}) to compute the average density of the DDD process with Eq.~(\ref{eq:ratchetpotential}).
Additionally, we note that the approximation to $O(R^2)$ in Eq.~(\ref{eq:ode_rho})
requires Taylor expansion of $Q(f_{\pm R}^{-1}(x))$ (near $x$) as defined in Eq.~(\ref{appeq:Q}). 
If we define $y=f_{\sigma}^{-1}(x)$, where $\sigma \in\{-R,R\}$, then we have
(where $Q=Q(x)$ when no argument is stated explicitly)
\begin{equation}
    Q(y)  \approx Q - \sigma QQ' + \sigma^2 \left[ Q(Q')^2 + \frac{1}{2!}Q^2Q'' \right] + O(\sigma^3)
\end{equation}
When the first and second terms are of the same order,
we expect the above Taylor expansion to fail. 
This gives the criterion $R<\frac{1}{\mathrm{max}(Q')}$.
 We show this criterion in Fig.~\ref{appfig:parameter_region} as a black solid line, 
 which agrees with the previous invertibility condition via the presence of a maxima.

In the above paragraphs, we have confirmed that the region of parameter space in which the truncation of Eq.~(\ref{appeq:multiple_options}) at $O(R^2)$ is valid when $P=100$.
We have additionally checked that the region of validity always decreases with increasing asymmetry ($P$) in the ratchet.
Finally, we note that the requirement of single-valued inversion of Eq.~(\ref{eq:local_split}), or equivalently the validity of Taylor expanding
to lowest order in $R$, means that a piecewise ratchet potential cannot be used when approximating Eq.~(\ref{eq:rho_full_sink}) by
Eq.(\ref{eq:ode_rho}) since no unique inverse exists, hence our use of the smooth ratchet potential of Eq.~(\ref{eq:ratchetpotential})
in this manuscript.

\section{Estimators of distributions measured in simulation}

For this section only, a `hat' over a variable indicates that it is a random variable, not an operator as in the main text. 
This random variable may be sampled from a time-dependent stochastic process. 
In this manuscript, $\{\hat{y}(t)\}_{\hat{m}(t)}$ will always represent the positions of
particles generated from Brownian dynamics simulations with division and death determining $\hat{m}(t)$ at each time step $t$.
We omit the time dependence of all random variables for brevity. 
The expectation of any function $g_{\hat{m}}(\{\hat{y}\}_{\hat{m}})$ of this set of random variables (at a specified time $t$) is calculated by
\begin{equation}
  \langle g_{\hat{m}}(\{\hat{y}\}_{\hat{m}})\rangle\equiv
  \sum_{\hat{m}=1}^{\infty}\int d\{\hat{y}\}_{\hat{m}} g_{\hat{m}}(\{\hat{y}\}_{\hat{m}})
  p_{\hat{m}}(\{\hat{y}\}_{\hat{m}},t).
\end{equation}
and an unbiased estimator of $\langle g_{\hat{m}}(\{\hat{y}\}_{\hat{m}}) \rangle$ is
\begin{equation}
  \bar{g}_{\hat{m}}(\{\hat{y}\}_{\hat{m}})
  =\frac{1}{N_s}\sum_{s=1}^{N_s}g_{\hat{m}_s}(\{\hat{y}_s\}_{\hat{m}_s})
\end{equation}
where the sum is over a set of length $N_s$ containing uncorrelated samples taken from simulation at time $t$. 
Typically in this paper, we take $N_s=10000$.

We now state the functions that, when averaged over,
are unbiased estimators of distribution functions such as $p_n^{(l)}$ and $\rho^{(l)}$.
The estimator, $\bar{p}_n$, of the full distribution for $n$ particles, is taken to be a sum over samples of the function
\begin{equation}
  p_{n,\hat{m}}(\{x\}_n,\{\hat{y}\}_{\hat{m}}) = \delta_{n,\hat{m}}
  \frac{1}{n!}\sum_{\pi\in S_n}
  \prod_{k=1}^n\delta(x_k-\hat{y}_{\pi(k)})
\end{equation}
where the sum is over all $\pi\in S_n$, where $S_n$ is the set of permutations in the symmetric
group on $n$ objects (this set is length $n!$). It follows from this equation that
\begin{align}
    &p_{n,\hat{m}}^{(l)}(x_1,\{\hat{y}\}_{\hat{m}})
    =\frac{\delta_{n,\hat{m}}(n-l)!}{n!}
    \sum_{i_1=1}^n\delta(x_1-\hat{y}_{i_1}) \nonumber\\
    &\times\sideset{}{'}\sum_{i_2=1}^n \delta(x_2-\hat{y}_{i_2})\cdots
    \sideset{}{'}\sum_{i_l=1}^n\delta(x_l-\hat{y}_{i_l}),
\end{align}
where the prime on the summation means excluding all previous indices (so the second sum is over $i_2\neq i_1$, the third sum is over $i_3\not\in \{i_1,i_2\}$, up to the $l$ sum which is over $i_l\not\in\{i_1,i_2,\dots,i_{l-1}\}$. 
Similarly, from Eq.~(\ref{eq:lbodydensity}) we have
\begin{align}
    &\rho^{(l)}(\{x\}_l,\{\hat{y}_{\hat{m}}\})
    = \sum_{i_1=1}^{\hat{m}}\delta(x_1-\hat{y}_{i_1}) \nonumber\\
    &\times\sideset{}{'}\sum_{i_2=1}^{\hat{m}}
    \delta(x_2-\hat{y}_{i_2}) \cdots
    \sideset{}{'}\sum_{i_l=1}^{\hat{m}}\delta(x_l-\hat{y}_{i_l}).
\end{align}

Note: simulation software is available through a custom package
which is run under the LAMMPS \cite{LAMMPS} software interface. This package is
available publically \cite{POP_package} under the github branch ``birthdeath\_package''.

\bibliographystyle{h-physrev}
\bibliography{bib}
\end{document}